\documentclass[preprint]{aastex}

\shorttitle{A Photometric Study of NGC 752}
\shortauthors{Anthony-Twarog, Deliyannis, Thomas, Twarog}

\begin{document}


\title{A $uvbyCa$H$\beta$ CCD Analysis of the Open Cluster Standard, NGC 752 \thanks{WIYN Open Cluster Study LXIX}}


\author{Bruce A. Twarog}
\affil{Department of Physics and Astronomy, University of Kansas, Lawrence, KS 66045-7582, USA}
\email{btwarog@ku.edu}

\author{Barbara J. Anthony-Twarog}
\affil{Department of Physics and Astronomy, University of Kansas, Lawrence, KS 66045-7582, USA}
\email{bjat@ku.edu}

\author{Constantine P. Deliyannis}
\affil{Department of Astronomy, Indiana University, Bloomington, IN 47405-7105 }
\email{cdeliyan@indiana.edu}

\and

\author{David T. Thomas}
\affil{Department of Physics and Astronomy, University of Kansas, Lawrence, KS 66045-7582, USA}
\email{wendysdbcb@outlook.com}




\begin{abstract}
Precision $uvbyCa$H$\beta$ photometry of the nearby old open cluster, NGC 752, is presented. The mosaic of CCD fields covers 
an area $\sim$42\arcmin\ on a side with internal precision at the 0.005 to 0.010 mag level for the majority of stars down to $V$ $\sim$
15. The CCD photometry is tied to the standard system using an extensive set of published photoelectric observations adopted as
secondary standards within the cluster. Multicolor indices are used to eliminate as nonmembers a large fraction of the low
probability proper-motion members near the faint end of the main sequence, while identifying 24 potential dwarf members between
$V$ = 15.0 and 16.5, eight of which have been noted before from Vilnius photometry. From 68 highly probable F dwarf members, 
we derive a
reddening estimate of $E(b-y)$ = 0.025 $\pm$ 0.003 ($E(B-V)$ = 0.034 $\pm$ 0.004), where the error includes the internal 
photometric uncertainty and the systematic error arising from the choice of the standard $(b-y$, H$\beta$) relation. With reddening
fixed, [Fe/H] is derived from the F dwarf members using both $m_1$ and $hk$, leading to [Fe/H] = -0.071 $\pm$ 0.014 (sem) and
-0.017 $\pm$ 0.008 (sem), respectively. Taking the internal precision and possible systematics in
the standard relations into account, [Fe/H] for NGC 752 becomes -0.03 $\pm$ 0.02. With the reddening and metallicity defined,
we use the Victoria-Regina isochrones on the Str\"omgren system and find an excellent match for $(m-M) = 8.30 \pm 0.05$ 
and an age of 1.45 $\pm$ 0.05 Gyr at the appropriate metallicity.

\end{abstract}


\section{Introduction}
Nearby star clusters are invaluable because they permit observational access to stars of lower luminosity 
with a specific age and composition using the full array of photometric and spectroscopic tools, 
often without the need for telescopes of large aperture. With the exception of the sparsely populated 
open cluster, Rup 147 \citep{CU13} at $(m-M)$ = 7.35, NGC 752 remains the nearest open cluster ($(m-M)$ = 8.4) 
older than 1 Gyr, with a well-determined age of 1.45 Gyr \citep{AT09}. While it is likely that the 
uncertainties in the fundamental cluster properties of Rup 147 will diminish as its membership is expanded 
and its individual stars are studied on a variety of photometric systems, the areal spread of over 
5 square degrees and the low density contrast have led to slow progress in the investigation of this intriguing 
object. By comparison, NGC 752 has a long history of photoelectric observations on virtually every major 
photometric system from broad-band UBVRI \citep{JO53, EG63, TA08} and Washington photometry \citep{CA86} 
to intermediate-band photometry on the extended Str\"omgren \citep{CB70, TW83, JO95, AT06}, 
Vilnius/Stromvil \citep{DP93, BA07, ZD10, BA11}, DDO \citep{JA79}, and Geneva \citep{RU81, RU88} systems. 
High and moderate-dispersion spectroscopy \citep{FJ93, FR02, SE04, CP11, RE12, NP13, MA13,BO15} is available 
for limited samples of both dwarfs and giants. The definitive proper-motion survey is \citet{PL91} (hereinafter PL)
and radial-velocity surveys include \citet{DA94, ME98, ME08, ME09}, but additional 
radial-velocity measures for 45 stars in the field of NGC 752 are also available in \citet{MA13}. 

Because of the magnitude and color range of the cluster color-magnitude diagram (CMD), NGC 752 should be 
an ideal candidate for standardization of any photometric system observed with a large format CCD on moderate to 
large aperture telescopes. In particular, because of its extensive list of photoelectrically studied stars 
on the  Str\"omgren system \citep{CB70, TW83, JO95} and, to a lesser extent, the $Caby$ system \citep{AT06}, 
NGC 752 has been regularly included in the standardization of our cluster work, most recently within our 
program to map Li evolution over a range of age and metallicity \citep{AT09, AT10, AT13, AT14, CU12, LB15} 
using open clusters with unusually well-defined fundamental parameters, critical for deriving precise, 
high-dispersion spectroscopic abundances.

However, despite the wealth of published data for NGC 752, as with Rup 147, its significant size of over a
square degree has limited the completeness of the CCD data available on any photometric system. To date,
the only significant CCD surveys within NGC 752 are the Vilnius studies of \citet{ZD10} and \citet{BA11}. The older study
covers 1.5 square degrees with a CCD camera of low spatial resolution (3.4\arcsec\ per pixel), while the
more recent analysis covers four overlapping fields of 12\arcmin\ $\times$ 12\arcmin\ each in the cluster core,
observed with the Vatican Advanced Technology Telescope with a CCD resolution of 0.37\arcsec\ per pixel. 
The purpose of this paper is to present precision CCD photometry of NGC 752 on the extended Str\"omgren system 
covering a field approximately 42\arcmin\  $\times$ 42\arcmin\ on the sky to below the current magnitude limit 
for complete proper-motion membership data. While the survey still covers only one-third of the potential area 
of the cluster, it does contain approximately two-thirds of the known members brighter than $V$ = 15.2, a magnitude 
range which may represent the actual faint limit for the main sequence members of this evaporating cluster \citep{FR89}. 
More specifically, the expanded standards will be adopted as primary contributors to the calibration of CCD data 
in two key clusters, NGC 7789 and NGC 2158, within our current cluster Li investigation.

The outline of the paper is as follows: Sec. 2 discusses the CCD observations and their reduction to the 
standard system for intermediate-band photometry; Sec. 3 uses the photometry, in conjunction with 
proper-motion membership, to identify and isolate probable cluster members which become the core data set 
for selecting single, main sequence stars for reddening, metallicity, age, and distance estimates in Sec. 4.
The extrapolated main sequence is also used to identify potential lower main sequence members of the cluster.  
Sec. 5 contains a summary of our conclusions.

\section{Observations and Data Reduction}

\subsection{Observations} 
Intermediate and narrow-band images of NGC 752 were obtained using the WIYN 0.9-m telescope during three
observing runs in Nov. 2010, Nov. 2012 and Dec. 2013. The cluster was observed as a primary source of standards
for the extended Str\"omgren and H$\beta$ systems, along with a range of field stars from the standard catalogs
for both systems. Because of the apparent brightness of the cluster stars and the wide range in color, NGC 752 
was observed on all usable nights - for zero-point and color calibration on photometric nights and for color-slope 
estimation on non-photometric nights. For the first two observing runs, the S2KB CCD was used at the $f$/7.5 focus 
of the telescope for a 20\arcmin\ $\times$ 20\arcmin\ field with 0.6\arcsec\ pixels. All seven filters were 
from the 3\arcsec $\times$ 3\arcsec\ filter set owned jointly by the University of Kansas and Mt. Laguna Observatory. 
For the Dec. 2013 run, the telescope was equipped with the Half-Degree-Imager (HDI), a 4K $\times$ 4K chip with 
a 29\arcmin $\times$ 29\arcmin\ field with 0.43\arcsec\ pixels. The seven filters were from the extended 
Str\"omgren set recently acquired for specific use with the HDI.

Bias frames and dome flats for each filter were obtained every night, with sky flats observed at twilight 
for the $u$, $v$, and $Ca$ filters when feasible for the S2KB runs; for the Dec. 2013 run, sky flats were obtained
every night for every filter in use each evening. Extinction stars and field star standards were observed
every photometric night for use with clusters without internal standards and as a check on the calibrations 
of cooler dwarfs within NGC 752 since the photoelectric data within the cluster are dominated by red giants 
and F dwarfs. As discussed earlier, while the cluster is an ideal calibration source for telescopes of small 
to modest aperture, its small distance modulus ($(m-M)$ $\sim$ 8.3) means that the areal coverage of the cluster 
field is large, making it a challenge to observe more than a handful of the internal secondary photoelectric 
standards in a single frame. To optimize coverage, NGC 752 was initially divided into four overlapping 
quadrants producing a composite field just over 33\arcmin\ on a side. For the Dec. 2013 run, the same 
quadrant centers were adopted, expanding the degree of overlap among the quadrants and expanding the 
areal coverage to $\sim$42\arcmin\ on a side. It should be emphasized that while the overlap among the 
frames can lead to dozens of measurements in each filter for stars near the cluster core, the exposure times 
for the frames were designed primarily with the internal photoelectric standards in mind, i.e. stars in the $V$ = 8 
to 12 range, so that precision photometry to $V$ = 16 is only feasible due to the large number of short exposures 
coupled to a limited set of longer frames designed to reach $V$ = 15 and below. This contrasts with our 
usual approach to cluster observations where emphasis is placed on minimizing photometric scatter for stars 
fainter than $V$ $\sim$ 16.5.

Standard IRAF routines were used to perform initial processing of the frames, i.e. bias-subtraction and flat-fielding.  
Illumination corrections were applied to frames obtained in 2012, but tests with the 2013 frames showed no statistically
significant difference between the photometry with or without correction, so none was applied. A fairly comprehensive 
discussion of our procedure for obtaining PSF-based instrumental magnitudes and merging multiple frames of a given 
filter can be found in \citet{AT00}. Instrumental magnitudes and indices were constructed separately for
the 2010/2012 data and for the 2013 data. The photometric indices from the S2KB chip and filters were transformed to the 
instrumental system defined by the HDI chip and its newer filters. The choice of fiducial instrumental system is arbitrary given
the ultimate need to transfer the instrumental indices to the standard system, but the larger field size, newer filters,
and greater overlap among mosaicked images made the HDI data the logical choice for defining the instrumental system. 
Prior to the predictable application of color-dependent corrections in transforming from one instrumental system 
(chip and filter set) to another, a comprehensive check was made for potential position-dependent offsets between the 
two data sets, a not improbable option given the nature of the overlapping CCD frames. No statistically significant 
discontinuities were discovered in magnitude or color at the overlap boundaries, lending credence to the validity of 
the frame-to-frame merger process. Position-dependent gradients across the composite S2KB field at the level of 0.05 mag 
were identified in the $y$ filter, the Str\"omgren equivalent for $V$, and eliminated using a quadratic in the X 
and Y positions; for color indices, trends with position were dramatically reduced or eliminated, but derived 
adjustments were applied whenever the residual patterns were found to be statistically significant.

Once transformed to a common instrumental system, the $y$ magnitudes and color indices from the two data sets were
combined using a weighting scheme tied to the inverse square of the derived standard errors of the mean for
the indivdual stars in both magnitude and color. 

\subsection{Internal Cluster Standards - Warmer Dwarfs and Red Giants}
Our standard procedure for calibrating CCD Str\"omgren photometry includes adoption of a single calibration 
equation for all colors and luminosity classes for $V$, H$\beta$, and $hk$, with the addition of a color term for $V$.
By contrast, for $m_1$, and $c_1$, separate equations with color terms are derived for warmer dwarfs ($b-y < 0.45$), 
for cooler ($b-y \geq 0.45$) dwarfs, and for cooler evolved stars; for $b-y$, a distinct calibration equation is 
sought and utilized for cooler dwarfs. The rich catalog of photoelectric observations 
within the cluster makes transformation of the blue dwarfs and red giants to the standard system straightforward. 
However, the photoelectric observations don't extend faint enough to include cool cluster dwarfs, a limitation which also
impacts the calibration of H$\beta$, a point we will return to below. 

For transformation to the standard system, the number and quality of the internal standards within NGC 752 are
strongly dependent upon which filter(s) are being transformed. For $V$, as of 1994 \citep{DA94} there were 5
sources of $V$-equivalent data; that number has since grown to 8. For $uvby$ and H$\beta$, the sources have 
doubled to 4 and 2, respectively, while for $Caby$, only one source exists \citep{AT06}, supplying $hk$ indices 
for 7 giants and 21 stars at the cluster turnoff. 

For the color indices $b-y$, $m_1$, $c_1$, and H$\beta$, and the $V$ magnitude, the photoelectric data from all 
available sources, unless otherwise noted, were merged after applying appropriate zero-point corrections and 
weighting by the inverse squares of the standard error of the mean for the individual data points.
Extensive discussion of an earlier attempt to merge the myriad photometric data sets for NGC 752, including Str\"omgren
photometry, can be found in \citet{DA94}, but the analyses of \citet{JO95} and \citet{TA08} emphasize a precision 
match to the zero-points for the fundamental $uvby$H$\beta$ and $V$ systems, so we will present a revised merger 
of each of the indices and $V$.

Prior to initiating the merger, a note on cluster identifications is necessary. As is common with well-studied 
open clusters, NGC 752 has accumulated an extensive list of numbering systems over the years, 12 as of this writing. 
The most comprehensive is that defined by WEBDA; the most recognizable in past discussions of the cluster is that 
of \citet{HE26}. When identifying stars in the discussions below, we will default to the \citet{HE26} (H) number, 
placing the WEBDA number in parentheses after the H number if they differ. Thus, if only the H number is supplied, 
it is assumed that the WEBDA and H numbers are the same. If a star has no H identification, only the WEBDA number 
will be supplied, preceded by W.

For H$\beta$ only two data sources are available, \citet{CB70} and \citet{JO95}. An increase of 0.008 mag was 
applied to the unevolved stars of \citet{CB70}, based upon a comparison of 12 turnoff stars common to the two data sets, and
the sets merged. \citet{CB70} also observed 4 red giants which have not been adjusted for the offset defined by the 
turnoff stars. The final list of 40 potential H$\beta$ standards ranges from 2.566 to 2.919.

For $b-y$, $m_1$ and $c_1$, we have data from \citet{CB70, TW83, JO95} and \citet{AT06}, though the last source supplies 
$m_1$ and $c_1$ solely for the 7 giants. For $c_1$, the systems of \citet{CB70} and \citet{JO95} are virtually 
identical; an offset of -0.001 was applied to the $c_1$ indices of \citet{CB70} and the data merged. From 17 
stars common with the merged sample, an offset of -0.010 $\pm$ 0.009 (sd) was found and applied to the 
$c_1$ data of \citet{TW83}. Only \citet{CB70} included red giants in their photometry and, while an offset 
of -0.010 in $c_1$ was also found, the overlap of 3 stars with \citet{AT06} was considered 
too small to provide a reliable offset, so the photometry was left unchanged. Lastly, one bright G star 
was observed as part of the catalog of \citet{OL93}. Since our cool-star $uvby$ standards are tied to this 
catalog, the star, H39, has been added to the standard list without change. All adjusted $c_1$ data were sorted and 
merged with weights tied to the inverse squares of the standard error of the mean for each star, leading to 
46 potential $c_1$ standards.

Identical procedures were followed for $m_1$ and $b-y$. No offset exists in $m_1$ or $b-y$ for the photometry 
of \citet{CB70} and \citet{JO95}. The derived adjustments from 17 stars for \citet{TW83} were +0.013 $\pm$ 0.011 (sd) 
and +0.002 $\pm$ 0.007 (sd) for $m_1$ and $b-y$, respectively. For \citet{AT06}, the $m_1$ indices for the giants 
remained unchanged but the $b-y$ indices for 20 dwarfs implied a correction of -0.009 $\pm$ 0.009 (sd). The 
one star from \citet{OL93} was added without adjustment. The adjusted indices were again merged and weighted 
by the inverse squares of the standard error of the mean, producing 46 potential $m_1$ standards and 47 
$b-y$ standards; H192 has only been observed by \citet{AT06} and H235 has been excluded as a variable.

For $V$, \citet{CB70} did not provide magnitudes. However, we attempted to merge the data of 
\citet{JO53, EG63, JH75, RU81, RU88, DP93,JO95, BA07}. The $V$ magnitudes of \citet{DP93} were eliminated 
after a number of stars exhibited unusually large residuals compared to multiple observations from other sources 
which showed no evidence for variability among these stars. Known variables H219 and H235 were excluded from 
the merger process.

For each source of photometry, residuals relative to \citet{TA08} were calculated and a color term was tested for 
statistical significance. The resulting coefficients, zero-points, and final residuals after adjustment 
are presented in Table 1. The revised magnitudes for all sources except \citet{EG63}, due to the larger 
than average scatter, were merged and combined. Stars for which the residual from the mean relation exceeded 
three sigma for the specific comparison were excluded from the derivation of the adjustment and from the final 
average for the star; these stars are identified by H number in the final column of Table 1. For 
\citet{RU81, RU88}, the standard error of the mean for the merger process was set at 0.010 
mag for all stars. The final merged database included 150 potential $V$ standards within the field of NGC 752.

Table 2 contains the final merged photoelectric photometry for all NGC 752 non-variable stars with any indices 
on the extended Str\"omgren and/or H$\beta$ systems. These data form the primary calibration standards within 
NGC 752. Stars with only $V$ magnitudes are not included within the Table.

\subsection{Internal Cluster Standards - Cool Dwarfs}
As noted earlier, none of the stars listed within Table 2 is classified as a cool dwarf. To allow 
transformation of the faintest cluster members to the standard system, the CCD frames for NGC 752 from
the Nov. 2010 run were treated as program frames and recalibrated with aperture photometry, with an emphasis on the
cooler dwarfs within the cluster field, i.e. dwarf stars too bright to be cluster members but bright enough to have 
adequate photometric precision to define the slopes and color terms for the indices of interest. This
particular run was chosen because it had the largest number of photometric nights with tie-ins to the 
cool dwarf field star standards.

Our calibrations to the standard extended Str\"omgren system are based on aperture photometry of field star standards
and of stars in NGC 752 for each photometric night.  For every frame contributing to the photometric calibration solution, 
aperture magnitudes for standard stars are obtained within apertures scaled to five times the full-width-half-maximum for 
the frame; sky annuli are uniformly chosen with the inner radius one pixel larger than the aperture and a uniform 
annular width. In addition to the stars of Table 2, a number of sources were consulted for field star standard 
indices, including the catalog of \citet{TA95} for $V$, $b-y$ and $hk$ indices, catalogs of $uvby$H$\beta$ observations by 
\citet{OL83,OL93,OL94}, and compilations of H$\beta$ indices by \citet{HM98} and \citet{SN89}. 

While our emphasis for the Nov. 2010 run was on the cool dwarfs, calibration relations were derived for all indices, 
colors, and luminosity classes. A single $b-y$ calibration equation was derived for warmer dwarfs and giant stars, with 
a separate calibration equation for dwarfs with $b-y \geq 0.45$. Calibration slopes and color terms for $m_1$ and $c_1$ 
for cooler giants were determined independently from those for bluer dwarfs or cooler main sequence stars, with the 
condition that the final relations mesh at the color boundary; the zero points for all relations were set by the 
standards of Table 2. The final step of extending the secondary standards within NGC 752 to cooler dwarfs used 
the average differences between the merged profile-fit photometry and indices from the entire run and the 
standard indices defined from the aperture photometry to transfer the PSF indices to the standard system. 
Since the shorter exposures in NGC 752 aren't exceptionally rich, crowding proved to be a non-issue and 
virtually every PSF star within the cluster frames could be translated to the standard system directly. The field star
calibration relations for the cool giants and hotter dwarfs from the Nov. 2010 run proved to be completely 
consistent with the values derived using the internal photoelectric standards as calibrators, thereby supplying 
secondary CCD standards from hotter dwarfs and red giants stars not already observed photoelectrically. Due to the
absence of cool dwarfs among the NGC 752 photoelectric standards, the secondary standards from the Nov. 2010 run 
for this group represent the primary, and potentially weakest, link to the standard system.   

\subsection{Variable Stars}
Before initiating the calibration of all the instrumental indices to the standard system, a check was made for 
potential variables, stars whose colors and, especially, magnitudes could produce distortions in the transformation 
relations. Such stars, if cluster members, could also prove useful in identifying and explaining spectroscopic 
and/or photometric anomalies compared to normal stars within the CMD. While the CCD frames weren't obtained 
with a variable star search in mind, the large degree of field overlap within the mosaic of frames has produced 
both high precision and, for the cluster core stars brighter than $V$ $\sim$ 15, often more than a dozen 
observations each in $b$ and $y$ extending over a three-year period. Therefore, the search for variables was 
carried out in the following way. The mean standard deviations for a single observation in $y$ and in $b$ were 
calculated as a function of $V$ for both the HDI and the S2KB photometry independently. Any star which exhibited 
photometric scatter in both $y$ and $b$ from the HDI data which was larger than 3.0 times the mean standard 
deviation at a given $V$ was checked within the S2KB data. If the star was within the S2KB field and exhibited 
similar scatter, it was tagged as a probable variable. Stars in the HDI field with excessive scatter for 
which the comparison in the S2KB data was inconclusive, as well as HDI variables located outside the S2KB field, 
were classed as possible variables. 

Our preliminary list contained 19 possible and 8 probable variables. As expected, the two well-studied cluster 
variables, H219 and H235, easily met all the criteria for probable variables. Each additional star was then checked for
potential signs of PSF contamination from a nearby star (closer than 4$\arcsec$) which could lead to apparent variability. 
This eliminated 2 probable variables and 5 possible variables, including one member, H182, a star we will see again
in Sec. 3.  Our final list totalled 6 probable variables and 14 possible variables brighter than $V$ = 17. Proper-motion
members among the probable variables are H58, H219, and H235, while possible variable members only include H129 and H156.
H58 and H129 are noted as potential variables in Table 2.

\subsection{Final Transformation Relations}
For all indices and the $V$ magnitude, a general calibration equation of the form

\noindent
INDEX$_{stand}$ = a$_{index}$*INDEX$_{instr}$ + b$_{index}*(b-y)_{instr}$ +  c$_{index}$

\noindent
was adopted. For $V$, H$\beta$, and $hk$, stars of all luminosity classes were treated initially as a group. For $b-y$, the
sample was separated into two groups, cooler dwarfs or blue dwarfs/red giants. The optimal transition color for the
separation was found to be $b-y$ = 0.46. For $m_1$ and $c_1$, calibration curves were attempted for three groups,
blue dwarfs, red dwarfs, and red giants. The red giant photoelectric standards were supplemented by the secondary 
CCD standards when necessary, while the red dwarf calibrations are based exclusively on the internal CCD standards.
The calibration coefficients, the number of stars used to define the relations, and the scatter among the residuals 
for each relation are presented in Table 3. 

Of the 150 potential standards for $V$ within the field of NGC 752, 116 fall within our CCD survey. Three of the
newly identified potential variables (H58, H129, H170) within the CCD field were removed, leaving 113 stars. A preliminary
calibration was tested and 8 additional stars were found to have larger than expected residuals. Once removed, 
the relation was rederived and the revised residuals tested for spatial variations across the CCD field using 
a quadratic function in both X and Y. It should be noted that the earlier merger between the S2KB and HDI data 
included spatially-dependent terms, but the adoption of the HDI frames as the true photometric system was partially  
arbitrary, leading to the need for a test of the HDI system with the photoelectric standards. The gradient 
in $V$ residuals across the 42$\arcmin$ field amounted to 0.02 mag, small but measurable given the precision 
of the instrumental magnitudes and the quality of the photoelectric standards. With application of a linear 
correction in X and Y, the final standard deviation in the residuals in $V$ from 105 calibration stars 
is $\pm$0.010 mag.  

Of the 8 stars with larger than expected residuals, the standard $V$ magnitude for H135 is based upon an average 
from 5 sources with a standard deviation among the measures consistent with no variability. However, close examination of the 
CCD data reveals a fainter star within 3$\arcsec$ of H135 which, if not excluded from the photoelectric flux, produces 
a $V$ magnitude too bright by 0.06 mag, enough to explain the discrepancy within the photoelectric data. A virtually 
identical result is found for H309; the presence of a star 2.55 mag fainter within 3$\arcsec$ of H309 leads to 
a photoelectric value too bright by 0.07 mag. H145 shows the largest discrepancy at 0.41 mag, with the sole 
observation from \citet{TA08} being too bright. With more than 2 dozen observations each in $b$ and $y$, our 
photometry for this star shows no indication of variability. As an independent check, the CCD $V$ mag from the 
Vilnius survey of \citet{BA11} for H145 is 12.72, 0.39 mag fainter than that of \citet{TA08}. A possibility is 
that the star observed by \citet{TA08} is H151, fainter and slightly redder than H145. A similar but less extreme 
discrepancy arises with H273. As noted earlier, comparison of the $V$ data of \citet{TA08} with that of 
\citet{RU81,RU88} led to elimination of this star from the merger of the latter data with the
standard list since the $V$ mag was too faint by 0.13 mag. Our data for this star exhibit the same offset 
while the \citet{BA11} CCD data show an offset of 0.12 mag. Based upon the $V$ mag and very red color, we conclude 
that the star listed as H273 in \citet{TA08} is possibly H272. H308 and H226 show no variability and have no nearby 
faint companions; the Vilnius $V$ magnitudes for these stars are in excellent agreement with ours, implying that 
the \citet{RU81, RU88} values are too bright by 0.15 and 0.07 mag, respectively. H100 is 9$\arcsec$ from H99, a star 1.3 mag 
brighter and an easy source of contamination for a photoelectric measurement; \citet{BA11} don't include this star in 
their survey. Finally, H176 is 0.04 mag fainter in our data than in \citet{RU81,RU88}. There are no stars with $V$ within 
5 mag of the star within 20$\arcsec$ of the star. The Vilnius $V$ mag for the star lies midway between the two 
discrepant measures, i.e. within 0.02 mag of our final result.
 
For H$\beta$, the standard deviation of the residuals among the 39 nonvariable photoelectric standards within the 
CCD field is $\pm$0.007 mag. Given the precision of the photometry, the small dispersion would appear to validate the 
adoption of a single calibration relation for stars in all luminosity classes. However, a preliminary analysis of the 
cluster using H$\beta$ defined by such a relation produced identical well-defined trends of increasing metallicity with 
decreasing H$\beta$ for main sequence members using either $hk$ or $m_1$ as the metallicity indicator. The similar 
trend from independently calibrated metallicity indices pointed to H$\beta$ as the common link to the problem, a 
result confirmed by a comparison to the metallicity obtained when $hk$ and $m_1$ are analyzed as a function of $b-y$ rather 
than H$\beta$. The underlying basis of the problem is the narrow temperature range which defines the stellar 
sample within the vertical turnoff of NGC 752. Excluding the single blue straggler, the range in H$\beta$ is only 0.08 mag. 
The blue straggler expands this range to 0.27 mag, but the inclusion of the four red giants
observed by \citet{CB70} extends and defines the relation among the cooler stars. While the 34 turnoff stars set the
definitive zero point for the H$\beta$ index, the trend of the residuals with H$\beta$ is set by the relative positions of the
sole blue straggler and the red giants. A relation excluding the blue straggler but retaining the giants with the
H$\beta$ of \citet{CB70} corrected for the offset defined by the unevolved stars produces a linear relation
with a statistically weak trend with H$\beta$, i.e. a slope near 1.0. If, as assumed in Table 2, the offset defined 
by the dwarfs doesn't apply to the red giants, the slope of the red giant calibration becomes 1.1. By contrast, 
dropping the giants but retaining the blue straggler creates a color dependence with a slope near 1.2. 

For stars classified as dwarfs at all colors, we have adopted the H$\beta$ relation defined by the 35 unevolved stars. For the
stars classified as giants, we will use the relation defined by the red giants without the offset \citep{JO95} included. 
We will return
to this point in Sec. 4 where the impact of the relation on both metallicity and reddening can be tested.   

For $hk$, all 27 photoelectric standards are within the cluster field. One star, H259, exhibited larger than expected 
residuals and was dropped from the calibration. The remaining 26 stars produced a scatter of $\pm$0.023, 
larger than the other Str\"omgren indices, but consistent with the significantly smaller number of photoelectric 
observations from only one source which were used to define the standard values. 

For $b-y$, from 46 red giant/blue dwarfs (H135 excluded) with photoelectric measures, the scatter among the residuals 
about the mean relation was $\pm$0.005 mag, the same scatter found for the cool dwarf relation defined by the 14 secondary 
CCD standards.

For $m_1$, to delineate separate calibrations for cool dwarfs, blue dwarfs, and red giants, the primary standards 
from Table 1 were supplemented by the 14 red giant and 14 cool dwarf secondary CCD standards. From the red
giant and blue dwarf relations, the 45 photoelectric standards of Table 1  (H135 excluded), exhibit a scatter about 
the mean relation of $\pm$0.009 mag. The secondary giant and dwarf standards show scatter of $\pm$0.017 mag and
$\pm$0.020, respectively, larger than for the photoelectric standards due to the more limited precision of the 
secondary CCD standards.

For $c_1$, separate calibrations were attempted for the cool dwarfs, cool giants, and blue dwarfs but, within 
precision of the secondary CCD standards dominating the cool dwarf sample, the relations derived for the dwarfs 
were found to be the same, so both samples were merged to obtain the common relation. For the red giant relation,
the 10 photoelectric standards of Table 2 were supplemented by 16 CCD secondary standards. One giant exhibited
larger than expected residuals and was dropped from the calibration. The final scatter about the 
mean relation for the 45 photoelectric standards (blue dwarfs and red giants) is $\pm$0.019 mag. For 
the 14 cool dwarfs, the scatter among the residuals is $\pm$0.029 mag, larger than for the photoelectric 
standards, as expected.

Before discussing the analysis of the photometry, particularly the probable members of NGC 752, it would be useful to
supply some insight into the classification of the cooler stars as giants or dwarfs. As explained above, this 
distinction is significant for the $m_1$ and $c_1$ indices and less so for $b-y$. For the study of clusters 
at greater distance, the dichotomy among cluster members is severely reduced since cooler dwarfs are often 
too faint to play an important role in the cluster analysis. For NGC 752, stars of approximately solar mass 
or lower have $V$ magnitudes near 13 and fainter, well within the limits of the current study. Equally important, 
proper-motion membership is limited to stars brighter than V $\sim$ 15.5, so identification of potential 
fainter members can be enhanced if field giants can be eliminated from the cooler sample.

For $m_1$ and $c_1$, the most common means of separating giants from dwarfs for cooler stars ($b-y$ \textgreater\ 0.5) has been
the use of two-color diagrams, either $m_1$ or $c_1$ versus $b-y$ (see, e.g. \citet{OL84}) or combinations of these indices
designed to enhance the separation, such as the LC parameter defined in \citet{TW07}. Two issues arise in the application of 
these approaches: (a) for the coolest stars, $b-y$ \textgreater\ 0.7, the separation of the indices between dwarfs and evolved
stars decreases with increasing color and (b) the luminosity distinction must be made using instrumental indices
prior to application of the often distinctly different calibration curves for each class. A practical alternative is
supplied by the $hk$ and H$\beta$ indices which traditionally do not require separate calibration curves based on luminosity class.
As detailed in \citet{TA95}, the $hk$ - $(b-y)$ relations for evolved and unevolved stars separate near $b-y$ $\sim$ 0.5,
with the separation increasing until $b-y$ $\sim$ 0.7, where the dwarf relation reverses and $hk$ declines with increasing
$b-y$, crossing the evolved star trend near $b-y$ = 0.8 but growing more disparate with increasing $b-y$. This pattern
is illustrated in Fig. 1, where stars classed as unevolved and evolved are plotted as open and filled circles,
respectively. The sample is composed of all stars with $V$ brighter than 16.0 with at least 3 observations in each filter and
final photometric errors below 0.015 in $b-y$ and 0.030 in $hk$. For the region near $b-y$ = 0.75 and beyond, 
a second criterion has been adopted to aid in distinguishing evolved stars from dwarfs. As detailed in \citet{TW07}, 
H$\beta$ loses sensitivity for cool giants, approaching a value near 2.55 defined by the relative width of the 
H$\beta$ filters. By contrast, for dwarfs, H$\beta$ continues to decline as $b-y$ increases, reaching a minimum 
near 2.48 before reversing the trend with increasing color. The pattern is illustrated in Fig. 2, where the stars 
are limited to those with errors in H$\beta$ below 0.020. The symbols have the same meaning as in Fig. 1. It should be
emphasized that while the final calibration for H$\beta$ has included a distinction between giants and dwarfs for
reasons noted earlier, use of a common relation for both classes leaves the pattern of Fig. 2 unchanged. The H$\beta$ scale becomes
slightly compressed with the giants hitting a limit at 2.56 instead of 2.55 and the dwarfs minimized at 2.50 instead of 2.485.

Some confusion of classification does occur, typically for the bluest stars for which the separation in Fig. 1 is least reliable and
photometric scatter can easily lead to misclassification. Fortunately, the differences in the final indices caused
by selecting the wrong calibration are also severely reduced in this color regime. For fainter stars without reliable
$hk$ or H$\beta$, the dwarf classification has been adopted by default.

Final photometry on the $uvbyCa$H$\beta$ system can be found in Table 4, where stars are sorted by magnitude and 
identified primarily by right ascension and declination (J2000.00), with CCD coordinates transferred to the 
system of \citet{HO00}. The sequential photometry columns are $V$, $b-y$, $m_1$, $c_1$, $hk$, and H$\beta$, followed by the
standard error of the mean in each index and the number of frames included for $y$, $b$, $v$, $u$, $Ca$, H$\beta$ narrow,
and H$\beta$ wide. The identification sequence is WEBDA(W), PL, \citet{HE26}(H), \citet{RV61}(RV), and \citet{ST85}(ST). The last
two columns are the membership probability from PL, if available, and the classification used to define the
calibration relations. Stars without membership data are assigned -1.  Photometry has been included only if a star has been observed
twice within each filter used to construct the magnitude/color index and the final errors in the magnitude/color index
fall below 0.050 for $V$ and H$\beta$, 0.070 for $b-y$, 0.075 for $m_1$ and $c_1$, and 0.10 for $hk$. A plot of the 
standard errors of the mean as a function of $V$ for $V$ and the five color
indices is shown in Fig. 3. At $V$ = 16, the typical standard errors of the mean are 0.010 mag for $b-y$, 0.015 mag
for $m_1$ and H$\beta$, and just under 0.020 mag for $hk$ and $c_1$.

\section{The Color-Magnitude Diagram}
The CMD based upon ($V, b-y$) for all stars in Table 4 is shown in Fig. 4. Stars for which the internal errors in $b-y$
are below 0.015 mag are plotted as open circles. While the probable location of the cluster CMD, particularly the giant
branch, is discernible, the majority of stars within the CCD field are definitely nonmembers, especially below $V$ = 15.
To reduce the confusion, a first cut is made based upon the proper-motion survey of PL. Of 175 stars with non-zero
membership probability, 124 lie within the CCD field. 

The second cut uses the radial-velocity measures of \citet{DA94, ME08, ME09, MA13}. Stars H177, H186, and H258, 
for which multiple radial-velocity measures showed no evidence for variability but a mean value significantly different 
from the cluster, were excluded as nonmembers. H64 is classed as a questionable member based upon radial-velocity measurements
\citep{ME09}; it has been excluded from the sample. The 120 remaining stars are plotted in Fig. 5. Open black circles are 
all dwarfs for which only proper-motion membership is available. Filled black circles are dwarfs with both proper-motion 
and radial-velocity membership, but too few radial velocities, usually one, to test for variability. One star with multiple 
radial-velocity measures but a large uncertainty in the final velocity due to rotational broadening of the spectral 
lines, H159, is plotted as a filled black circle. Filled red circles (triangles) are stars where multiple radial-velocity 
measures are consistent with membership and single-star status. Filled blue circles (triangles) are stars where 
multiple radial-velocity measures are consistent with membership and binarity. Two of the stars classed as radial-velocity 
nonmembers by \citet{MA13} based upon a single velocity measurement (P552, P828) are known single-lined spectroscopic 
dwarfs and are plotted as such. Four dwarfs have single radial-velocity measures from \citet{MA13} which deviate 
significantly from the cluster mean, implying they could be nonmembers and/or binaries. These are plotted as green stars. 
P964 (H244) is a radial-velocity member with no evidence for binarity according to \citet{DA94} and \citet{ME09}; 
\citet{MA13} identify it as a nonmember based upon a single deviant measure. We have plotted it as a single-star member. 

While the cluster CMD is now extremely well-delineated, deficiencies in our categorization of the stars still remain. The
stars for which only proper-motion membership is available fall into two categories. At the faint end, where the membership
probabilities approach single digits, the four bluest stars between $V$ = 14 and 15.5 are highly probable nonmembers,
background dwarfs whose proper-motion uncertainties place them marginally within the range of the cluster. Another six
appear to lie on the main sequence, while a seventh would need to be a binary based upon the position in the CMD. We will return
to these below. Surprisingly, the second group of stars with minimal information includes 8 stars which populate the red hook
at the top of the main sequence at $V$ = 10.5 or brighter. Because of their apparent brightness relative to the field, 
it is almost certain that these stars would be confirmed as members if radial-velocity data were available. 
Their importance lies in the fact that almost all stars bracketing this region of the CMD are classed as binaries.
If single stars, given that these are the approximate progenitors of the red giants, a more comprehensive investigation 
of their physical state could shed significant light on the confusing mix of abundances for Li and CNO found among the giants
\citep{AT09, BO15}.

Moving down the main sequence, with the exception of the highly probable faint nonmembers, three stars appear to sit 
slightly blueward of the main sequence. The single-star member near $V$ = 12.3 is H182, originally tagged as a potential
variable from the scatter in the photometry but now believed to suffer from contamination by a nearby fainter star, thus
explaining its odd position in the CMD. Two low-probability members, W6341 (proper-motion membership probability = 2\%) 
and W7384 (1\%), without radial-velocity information at $V$ $\sim$ 14.75 comprise the rest of the probable blue CMD deviants. 
Three of the four stars plotted as starred points due to a deviant single radial-velocity measure \citep{MA13} fall above 
the main sequence. One of these (H90, 23\%) appears to be too bright to be a binary member of the cluster and is 
photometrically classed as a giant. A similar classification befalls the brightest of the green stars, making both of 
these highly likely nonmembers. H156 (87\%) is classed as dwarf but has a position in the CMD which is marginally too 
bright to be within the cluster. Only the faintest green star is both a dwarf and within the confines of the cluster CMD. 
Finally, the reddest star in the plot at $V$ $\sim$ 12.4 is classed photometrically as a field red giant.

As an independent means of identifying potential nonmembers through an anomalous CMD position, we turn to the $V$ - $hk$ diagram 
of Fig. 6. $hk$ has the advantage of being more sensitive to changes in temperature than $b-y$ while remaining relatively 
insensitive to reddening. Equally important, increased reddening in $b-y$ moves $hk$ toward lower/bluer values; 
$Ehk$ = -0.16*$E(b-y)$. Symbols in Fig. 6 carry the same meaning as in Fig. 5; the four stars with $b-y$ less than 0.50
located well below the main sequence of Figure 5 have been excluded from the figure and will not be discussed further. 
In Fig. 6, 2 stars immediately stand out by lying below the main sequence near $V$ = 14.75; these two stars are W6341 and W7384, 
also tagged for lying blueward of the main sequence in Fig. 5. All other stars have positions on or above the 
main sequence, but a closer analysis reveals that while the sequence of points with increasing $b-y$ should map to a similar
sequence with increasing $hk$, some stars have $hk$ indices which are inconsistent with their order in $b-y$. A trivial example
comes from the reddest star on the lower main sequence, W496 (1\%), located in a position in Fig. 5 which would imply that it must be
a binary. However, in Fig. 6 this star is located at a much bluer color, placing it on the expected single-star main sequence. Equally 
anomalous are two filled circles (H69, H159) in the turnoff region which lie significantly redward of the main sequence hook, 
almost within the subgiant region, while remaining blueward of this boundary in Fig. 5.

The color sequence discrepancies are readily apparent in Fig. 7, where $b-y$ is plotted relative to $hk$ for unevolved stars. 
A well-defined trend delineates the cluster main sequence, but a few points stand out. The blue point above the relation at 
$b-y$ = 0.34 is H235, an eclipsing binary with a significant scatter in all indices. The three discrepant points redward of
the relation between $hk$ = 0.55 and 0.65 are H182, H69, and H159. H182 is the likely contaminated star which was too blue
in $b-y$; making $b-y$ too small should lead to $hk$ being too large, as observed. As noted earlier, H69 is a radial-velocity 
member based on only one measurement \citep{ME09}, making it impossible to determine if this star is a possible binary. 
H159 has a radial velocity consistent with membership from multiple observations, but the uncertainty in the final value is 
indeterminate due to the apparently high rotation rate for the star. Its position well above the red hook would make a binary 
nature highly plausible, but its position could be tied to its high rotation rate, following the arguments of \citet{BH15} that
rapid rotators at a given age are longer-lived stars of higher mass which should appear brighter than slower rotators. 
Moving toward the redder stars, the single-lined spectroscopic binary H313 is the deviant point above the relation near 
$hk$ = 0.7. The deviant open circle at $hk$ = 1.0 is the previously tagged, low probability member W6341. 
The anomalous position of the projected lower main sequence binary now becomes apparent at $hk$ = 1.3. Finally, the green, 
starred points all lie above the mean relation, indicating they are too blue in $hk$ for their $b-y$ colors, adding credence 
to the radial-velocity classification as nonmembers. Only the reddest of the starred points is marginally consistent with
photometric membership.

\section{Cluster Parameters}
\subsection{Reddening and Metallicity}
Given the extensive and growing literature on NGC 752, the primary purpose of this study is not a comprehensive rediscussion 
of the cluster reddening and metallicity. However, a strength of the extended Str\"omgren system is the ability to supply 
precise reddening and metallicity for individual stars, the latter from two independent indices, $m_1$ and $hk$, with optimal 
sensitivity for F dwarfs, the exact range covered by the stars at the turnoff of NGC 752 and extending to $V$ $\sim$ 13. 
Fig. 8 illustrates the pattern between $b-y$ and H$\beta$ for all possible proper-motion members, excluding the faint blue 
probable nonmembers below the main sequence in Fig. 5 and the one 
blue straggler at (H$\beta$, $b-y$) = (2.931, 0.016). Since H$\beta$ is reddening independent, the vertical scatter 
in $b-y$ has two dominant sources, photometric error and reddening. Among the brighter stars, excluding a handful of 
deviant points, the standard deviation in $b-y$ at a given H$\beta$ is below $\pm$0.011 mag. A photometric scatter in H$\beta$ 
of $\pm$0.008 alone would explain this. Note also that there is no evidence for separation of the sample into 
binaries and single stars so elimination of binaries would reduce the final precision without altering the cluster mean. 
We conclude that there is no evidence for reddening variation across the cluster face, not surprising for a cluster with $(m-M)$ = 8.4.

Among the hotter stars, the two most deviant points are the usual suspects, the eclipsing binary H235 (filled blue circle) on 
the high $b-y$ side and H182 (filled red circle) on the low $b-y$ side. Of the seven open circles (only proper-motion 
membership available) three fall on the low side of the mean relation. Whether or not this is significant must await 
spectroscopic observations of these stars.

At the redder end of the sample, the cluster red giants emerge, following the same pattern laid out in Fig. 2, with a lower
limit of 2.55 to H$\beta$ at increasing $b-y$. Among the potential lower main sequence stars, there is an apparent 
shift to larger $b-y$ among many of the stars with H$\beta$ less than 2.57. Based upon the pattern illustrated in Fig. 2 
and the location of the cluster red giants in this figure, we conclude that many of the stars with low proper-motion 
membership at the base of the main sequence are evolved background stars. This conclusion has already been confirmed 
for the two bluest starred points. Among the deviants points sitting between the dwarfs and the giants is W6341 again, 
making it an anomaly in every diagram studied and a likely evolved background field star.

As we have done consistently in our cluster work, we will derive the reddening from two Str\"omgren relations from \citet{OL88} and
\citet{NI88}, a slightly modified version of the original relations derived by \citet{CR75, CR79}. Reddening estimates are derived 
in an iterative fashion. The indices are corrected using an initial guess at
the cluster reddening and the intrinsic $b-y$ is derived using the reddening-free H$\beta$ adjusted for metallicity and evolutionary
state. A new reddening is derived by comparing the observed and intrinsic colors and the procedure repeated. The reddening
estimate invariably converges after 2-3 iterations. To derive the reddening, one needs to correct $b-y$ for metallicity, so
a fixed [Fe/H] is adopted for the cluster and the reddening derived under a range of [Fe/H] assumptions that bracket the final
value. The complementary procedure is to vary the mean reddening value for the cluster and derive the mean [Fe/H]. Ultimately,
only one combination of $E(b-y)$ and [Fe/H] will be consistent.

The sample for metallicity and reddening estimation consists of 68 stars out of an original sample of 74. We have eliminated 
the deviants identified in Fig. 7, as well as H193 and the variable H219. H193 is classified
as an Fm star \citep{GA72}, leading to [Fe/H] = +0.7 from $m_1$. [Fe/H] from $hk$ is consistent with the cluster mean, but it will
be excluded from both averages. Two additional stars classified as possible Fm stars, H58 and H234, generate metallicities
consistent with the cluster mean and have been retained. For NGC 752, the metallicity defined by $m_1$ was varied between 
[Fe/H] = -0.30 and +0.30, generating a range of $E(b-y)$ = 0.030 to 0.020 for the relation of \citet{NI88} and 0.030 to 
0.016 for \citet{OL88} from 68 stars; in all cases, the standard deviation of
a single measurement is only $\pm$ 0.011 mag. The slightly higher reddening for F stars using 
the \citet{NI88} relation compared to that of \citet{OL88} is a consistent occurrence from such comparisons \citep{CA11, AT14}.

With $E(b-y)$ set at a range of values between 0.020 and 0.030, [Fe/H] has been derived from both $m_1$ and $hk$, using H$\beta$ 
as the primary color index. The best fit for simultaneous reddening and metallicity from $m_1$ produces $E(b-y)$ = 0.025 $\pm$ 0.003, 
where the uncertaintly is totally dominated by the systematics of the intrinsic color relations and based upon one-half the difference 
between the two values. If $E(B-V)$ = 0.73*$E(b-y)$, the reddening estimate from Str\"omgren data alone is $E(B-V) = 0.034 \pm 0.004$.
The final [Fe/H] from 68 stars is -0.071 $\pm$ 0.014 (sem) and -0.017 $\pm$ 0.008 (sem) from $m_1$ and
$hk$, respectively. In contrast with our preliminary analysis using a common  slope for calibration of H$\beta$, a plot of 
the individual abundances as a function of H$\beta$ in Fig. 9 shows no trend with decreasing H$\beta$ for either $m_1$ or $hk$. 
The standard deviation in [Fe/H] from $hk$ is noticeably smaller than that from $m_1$, a reflection of the 
greater metallicity sensitivity of $hk$ over $m_1$ for F dwarfs. By contrast, the dispersion becomes comparable for the 
two indices near the boundary of the early G stars, illustrating that at cooler temperatures, the $m_1$ and $hk$ indices undergo
significant increases with decreasing temperature, making the metallicity indices very sensitive to small changes in H$\beta$, 
and generally less reliable indicators of [Fe/H] \citep{TA95, AT02, TW07}.

Both the reddening estimate and the metallicity are in excellent agreement with the most recent work on the cluster. The definitive
discussion of the reddening is that of \citet{TA07} where the adopted value becomes $E(B-V)$ = 0.044 $\pm$ 0.003. 
The most recent high-dispersion spectroscopy by \citet{BO15} of 10 red giants leads to [Fe/H] = -0.02 $\pm$ 0.05 while 
\citet{MA13} derive [Fe/H] = -0.063 $\pm$ 0.013 from 36 main sequence stars, adopting $E(B-V)$ = 0.035. More complete 
discussions of all previous spectroscopic work can be found in these papers.

\subsection{Extending the Main Sequence}
While the proximity of a star cluster theoretically supplies access to a homogeneous sample of low mass stars, for objects like NGC 752
and Rup 147, the greater age and low stellar density can mean that the few stars which remain bound to the cluster can easily be lost
in the confusion of a rich background field. As demonstrated in previous sections, many of the lower probability, proper-motion members
fainter than $V$ = 14 have indices and CMD positions incompatible with membership in NGC 752. Identification of photometric members
fainter than $V$ = 14 has been attempted by \citet{BA11} using Stromvil indices and CMD location. Between $V$ = 15 and 16.5, they 
identify 12 potential photometric members. 

While the precision of the indices varies as a function of location in the field due to the partially overlapping
mosaic of frames, the $V, (b-y)$ combination remains 
reliable for almost all stars down to at least $V$ = 16.5. In an effort to identify
potential cluster members, we have selected all stars with $V$ between 15.0 and 16.5 and $b-y$ greater than 0.5 
with photometric errors in $b-y$, H$\beta$ and
$hk$ below 0.015, 0.020, and 0.030 mag, respectively. These stars were plotted in the CMD and the cluster main sequence from
known members above $V$ = 15 was extended. Any star which deviated from the probable main sequence by more than 0.020 mag in 
$b-y$ on the blue side was eliminated. On the red side, stars were kept if they fell within 0.6 mag in $V$ of the projected 
unevolved main sequence. Stars were then checked to see if their indices classified them as dwarfs or evolved stars.
Of the 12 stars tagged as photometric members by \citet{BA11}, eight met all of our precision criteria. Of these, six are confirmed
as probable members (W6878, W6932, W6962, W7010, W7311, W7578), while two are excluded (W7176, W7390). The remaining four
were initially excluded due to larger errors in $hk$ and/or H$\beta$. Of these, from $V, (b-y)$ alone, W7274 is a probable
nonmember, while W7039, W7187, and W7346 are consistent with main sequence stars.  W7039 presents a special case. Its $hk,(b-y)$
position in Fig. 1 clearly places it among the dwarf sample. However, its location in the H$\beta$ - $(b-y)$ plot (Fig. 2) 
classified it as a giant, so it was calibrated as a giant. If we require it to be a dwarf and apply the dwarf calibration for $b-y$, 
the final color shifts 0.04 mag to the blue, making this star inconsistent with cluster membership. It has been eliminated from
the membership discussion. From outside the survey area of \citet{BA11} 16 stars have been identified as probable members; their WEBDA 
identifications are 6415, 6476, 6516, 6528, 6559, 6628, 6766, 6795, 6845, 6891, 7022, 7132, 7156, 7267, 7537, and 7669. We emphasize
again that some of these stars are located above the unevolved main sequence in a position indicative of binarity, if they are 
members. The CMD of the selection is shown in Fig. 10. Symbols have the same meaning as in Fig. 5, with the following additions:
open black squares are the 16 probable members from our data alone, open blue squares are the members in common with \citet{BA11}, 
blue crosses are the photometric members of \citet{BA11} which we classify as nonmembers, red crosses are the stars identified
as nonmembers from our data alone, and green crosses are stars which lie on/near the main sequence but have indices which 
classify them as evolved stars. 

\subsection{Distance and Age}
One of the rare sets of available isochrones which include models transformed to the Str\"omgren system is the 
Victoria-Regina (VR) set of isochrones \citep{VA06}. Fig. 11 shows the scaled-solar models for [Fe/H] = -0.04,
ages 1.3, 1.4, and 1.5 Gyr, adjusted for $E(b-y)$ = 0.025 and $(m-M)$ = 8.30. All stars classed as binaries and/or
likely nonmembers, as discussed in previous sections, have been eliminated. Symbols have the same meaning as in
Figs. 5 and 10. Keeping in mind that no membership information beyond location in
the CMD is available for the square points, the isochrones supply an exceptional match to the entire 
CMD with an adopted age of 1.45 $\pm$ 0.05 Gyr.
For comparison, the analysis of \citet{AT09} using the composite broad-band $BV$ data of \citet{DA94} and the 
\citet{DE04} ($Y^2$) isochrones with [Fe/H] = -0.05 and $E(B-V)$ = 0.035 derived an age of 1.45 Gyr and $(m-M)$
= 8.4. The slightly smaller modulus is within the uncertainties of the broad-band result.
\citet{BA11} use Vilnius photometry to derive $E(B-V)$ = 0.048 and [Fe/H] = -0.16; from comparison to older Padova 
isochrones \citep{BR93, FA94}, they find an age of 1.4 Gyr and an apparent modulus of $(m-M)$ = 8.38.

\section{Summary and Conclusions}
With increasingly larger CCD cameras becoming the standard on telescopes of medium and large aperture, a challenge for
many less traditional photometric systems is calibration to a standard system. Unlike the extensive networks of
standards used to define all-sky photometry for broad-band systems, intermediate and narrow-band standards are often
limited to brighter and/or single stars within the field. While there are thousands of stars observed on the extended
Str\"omgren system, outside of star clusters these are predominantly brighter than $V$ $\sim$ 9.5. As an initial step
to remedy this problem, a 42\arcmin\ x 42\arcmin\ field centered on NGC 752 has been observed to $V$ = 16.5, with 
precision better than 0.010 mag for all indices for the majority of stars to $V$ $\sim$ 15.

The proximity of NGC 752 ensures that even with a telescope of modest size it is possible to reach stars of late K spectral
type, fainter than the current limit of radial-velocity and proper-motion surveys of the cluster. Using multicolor 
indices, membership has been tested for the faintest stars with nonzero proper-motion probabilities from PL. 
As expected, a majority are identified as likely nonmembers. Fortunately, the precision of the data allows
us to extend the probable main sequence to $V$ = 16.5, identifying 24 possible members, single and binary, between
$V$ = 15 and 16.5. The 12 photometric members identified by \citet{BA11} within the cluster core have been 
checked and and we concur on likely membership for 8 of them. The 16 additional photometric members are outside
the area of the Vilnius study.

Using the photometry contained within this sample alone, the fundamental parameters for NGC 752 are rederived. 
The cluster is found to have $E(b-y)$ = 0.025 $\pm$ 0.003 ($E(B-V)$ = 0.034 $\pm$ 0.004) and [Fe/H] = -0.03 $\pm$ 0.02 
from 68 F stars, using a weighted average from $m_1$ and $hk$. With a well-defined main sequence of predominantly single 
stars, the Str\"omgren CMD
is exceptionally well matched by an age of 1.45 $\pm$ 0.05 and $(m-M)$ = 8.30 for [Fe/H] = -0.04, in excellent
agreement with recent work tied to broad-band data. 

Despite the apparent brightness of the stars near the turnoff, some questions remain regarding membership and
binarity for stars near and above the main sequence red hook, the stars which populate the red giant branch and
clump. Two stars (H69, H159) with anomalous $hk$ values could be either nonmembers or stars demonstrating the 
effects of rapid rotation, while the supposedly single star, H108, lies well above the hook where no normal star
should be found. Finally, the one definitive blue straggler, H209, has never exhibited radial-velocity or
photometric variability; its age from comparison to the VR isochrones is less than 0.1 Gyr.

\acknowledgments
The paper has been significantly improved by the valuable and thoughtful comments of an anonymous referee who justifiably 
insisted that we probe deeper into the metallicity trends identified in an earlier version of the manuscript.
Extensive use was made of the WEBDA database maintained by E. Paunzen at the University of Vienna, Austria 
(http://www.univie.ac.at/webda). The filters used in the program were obtained by BJAT and BAT through NSF 
grant AST-0321247 to the University of Kansas. NSF support for this project was provided to BJAT and BAT 
through NSF grant AST-1211621, and to CPD through NSF grant AST-1211699.

\clearpage

\begin{figure}
\includegraphics[width=\columnwidth,angle=0,scale=0.80]{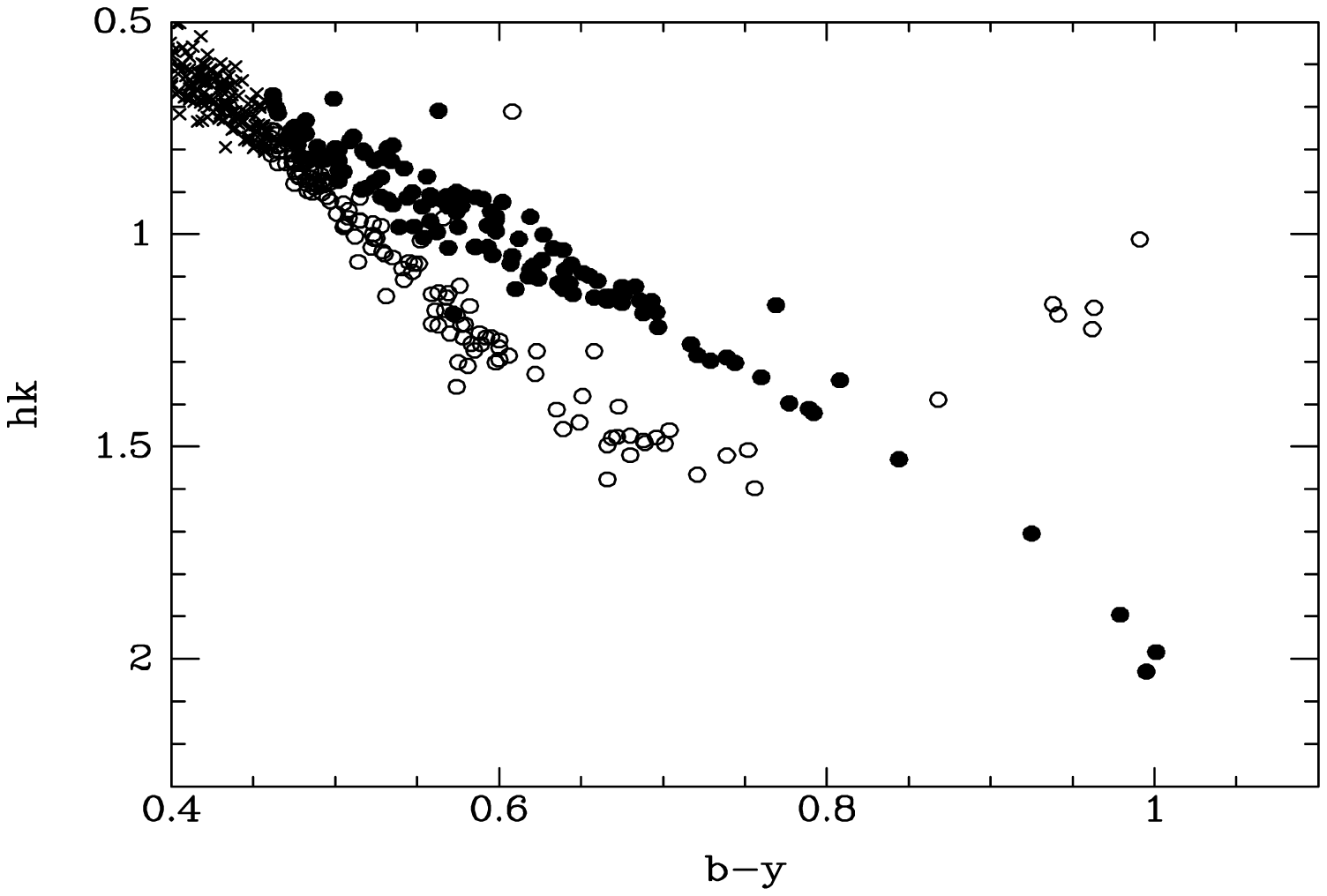}
\caption{Using $hk$ - $(b-y)$ to separate stars by luminosity class: stars classed as unevolved and evolved are 
plotted as open and filled circles, respectively. Crosses are stars for which no classification is possible.}
\end{figure}
\clearpage

\clearpage
\begin{figure}
\includegraphics[width=\columnwidth,angle=0, scale=0.80]{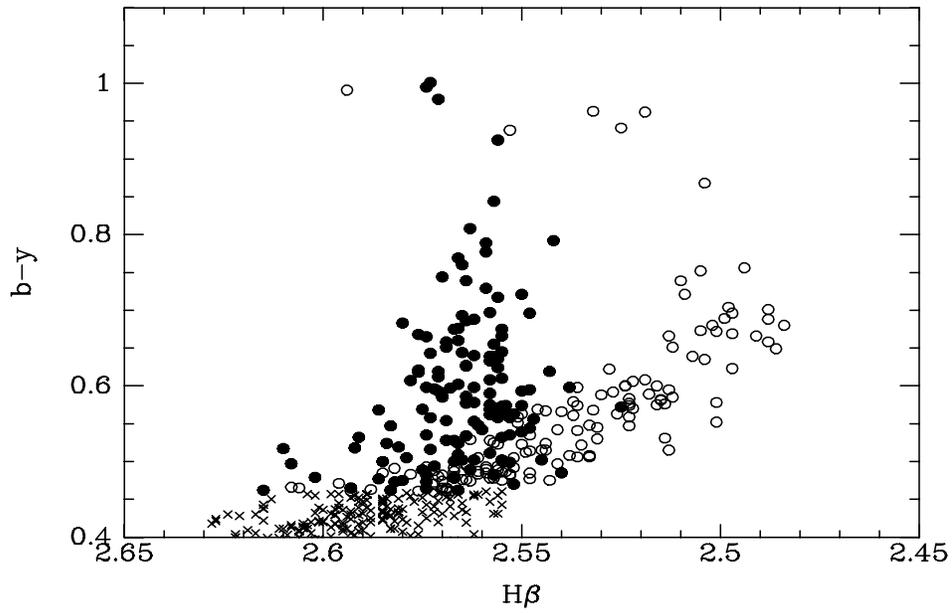}
\caption{Same symbols as Figure 1, using $b-y$ versus H$\beta$ to separate stars by luminosity.}
\end{figure}

\begin{figure}
\includegraphics[width=\columnwidth,angle=270,scale=0.80]{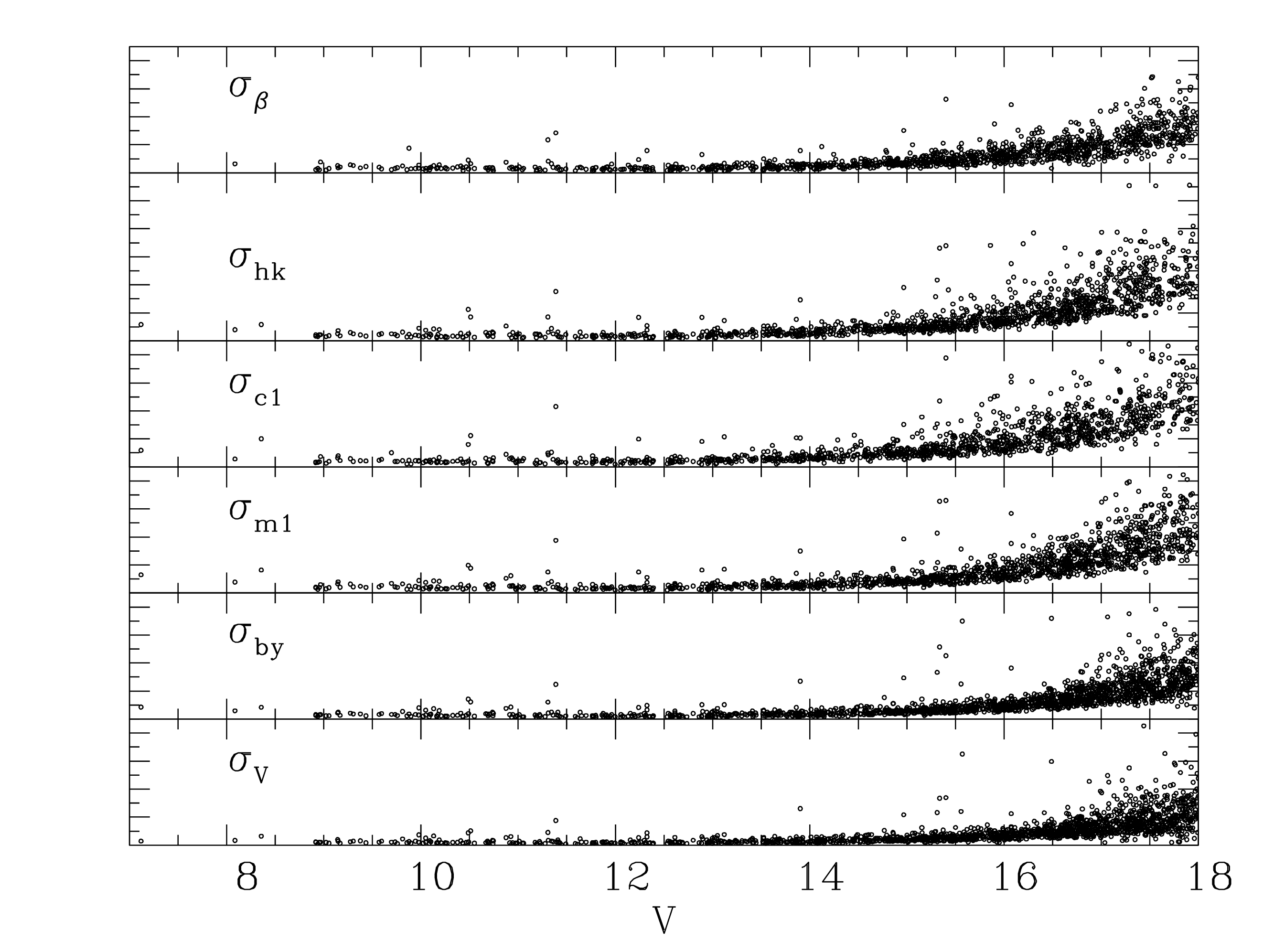}
\caption{Average photometric errors in $V$, $(b-y)$, $m_1$, $c_1$, $hk$ and $H\beta$ as a function of the $V$ magnitude. 
The panel heights are scaled proportionately to the physical range, with major tick-marks indicative of 0.02 mag.}
\end{figure}

\begin{figure}
\includegraphics[width=\columnwidth,angle=0, scale=0.80]{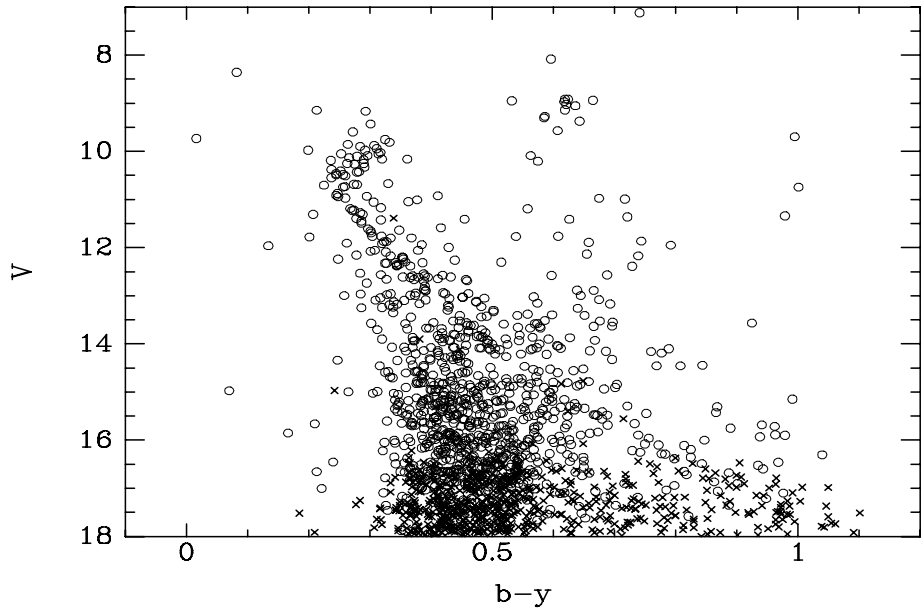}
\caption{CMD for all stars in Table 4. Open circles are stars with photometric errors in $b-y$ $\leq$ 0.015. Crosses are
stars with errors in $b-y$ larger than 0.015.}
\end{figure}

\clearpage
\begin{figure}
\includegraphics[width=\columnwidth,angle=270, scale=0.80]{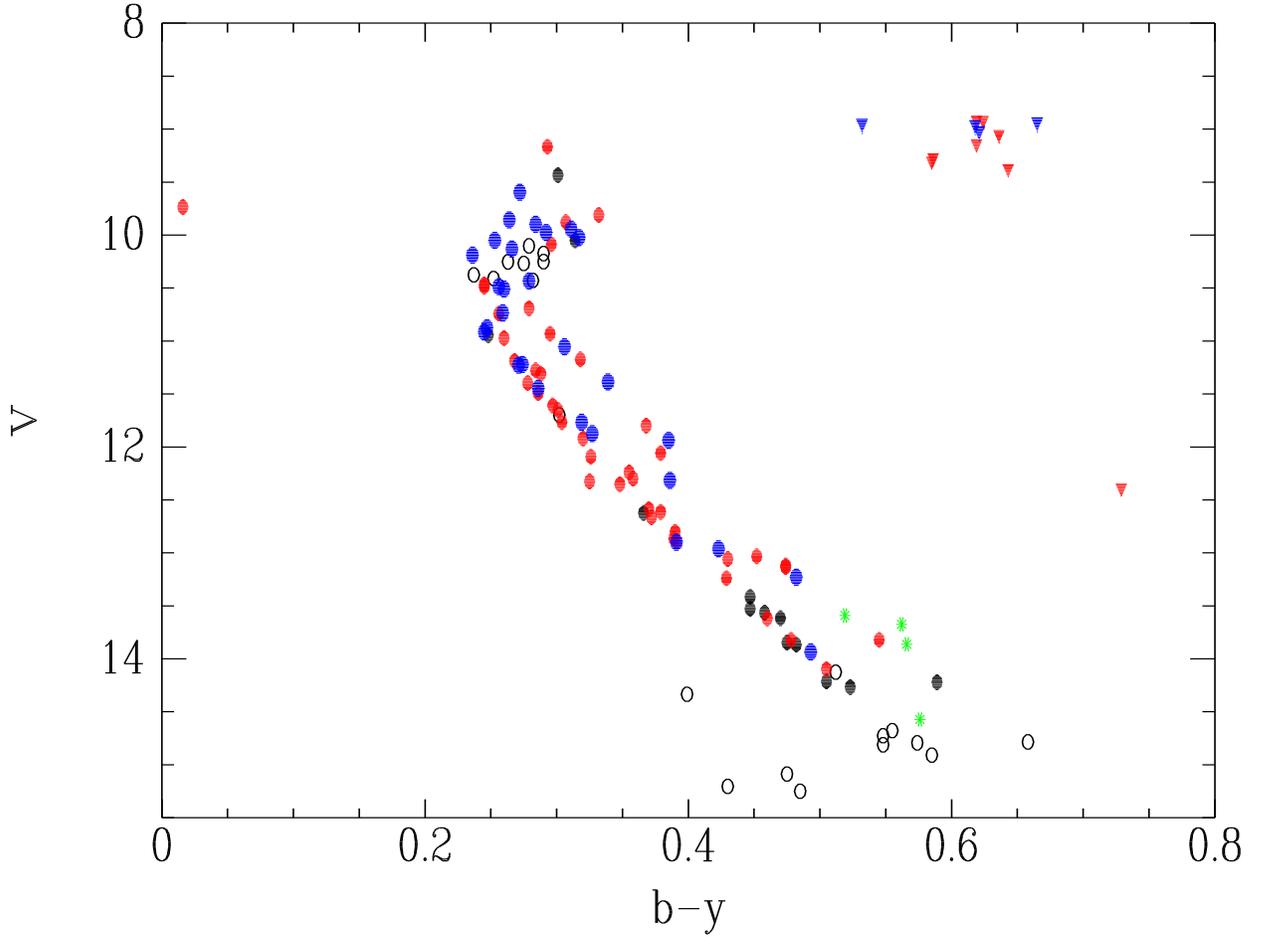}
\caption{CMD of all stars with nonzero proper-motion membership probabilities. Open black circles are stars without radial 
velocities. Filled black circles are stars with radial-velocity membership, but too few to determine binarity. Filled red 
circles (triangles) are dwarf (giant) members without radial-velocity evidence for binarity. Filled blue circles (triangles) 
are dwarfs (giants) classed as member binaries. Four dwarfs with single radial-velocity measures which deviate significantly 
from the cluster mean, implying they could be nonmembers and/or binaries, are plotted as green stars.}
\end{figure}

\clearpage
\begin{figure}
\includegraphics[width=\columnwidth,angle=270, scale=0.80]{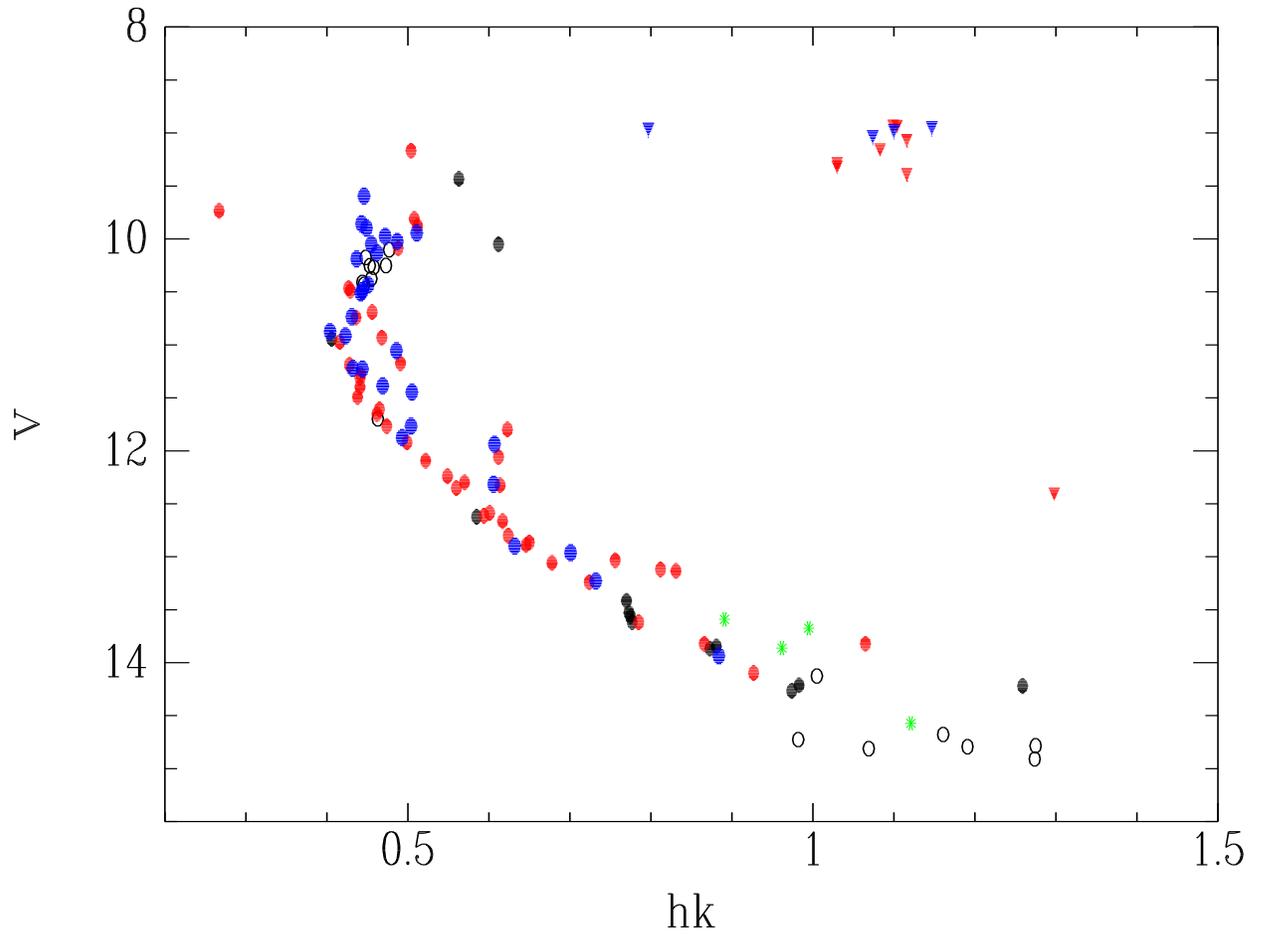}
\caption{CMD based upon $hk$ as the temperature index. Symbols have the same meaning as in Fig. 5.}
\end{figure}

\clearpage
\begin{figure}
\includegraphics[width=\columnwidth,angle=270, scale=0.80]{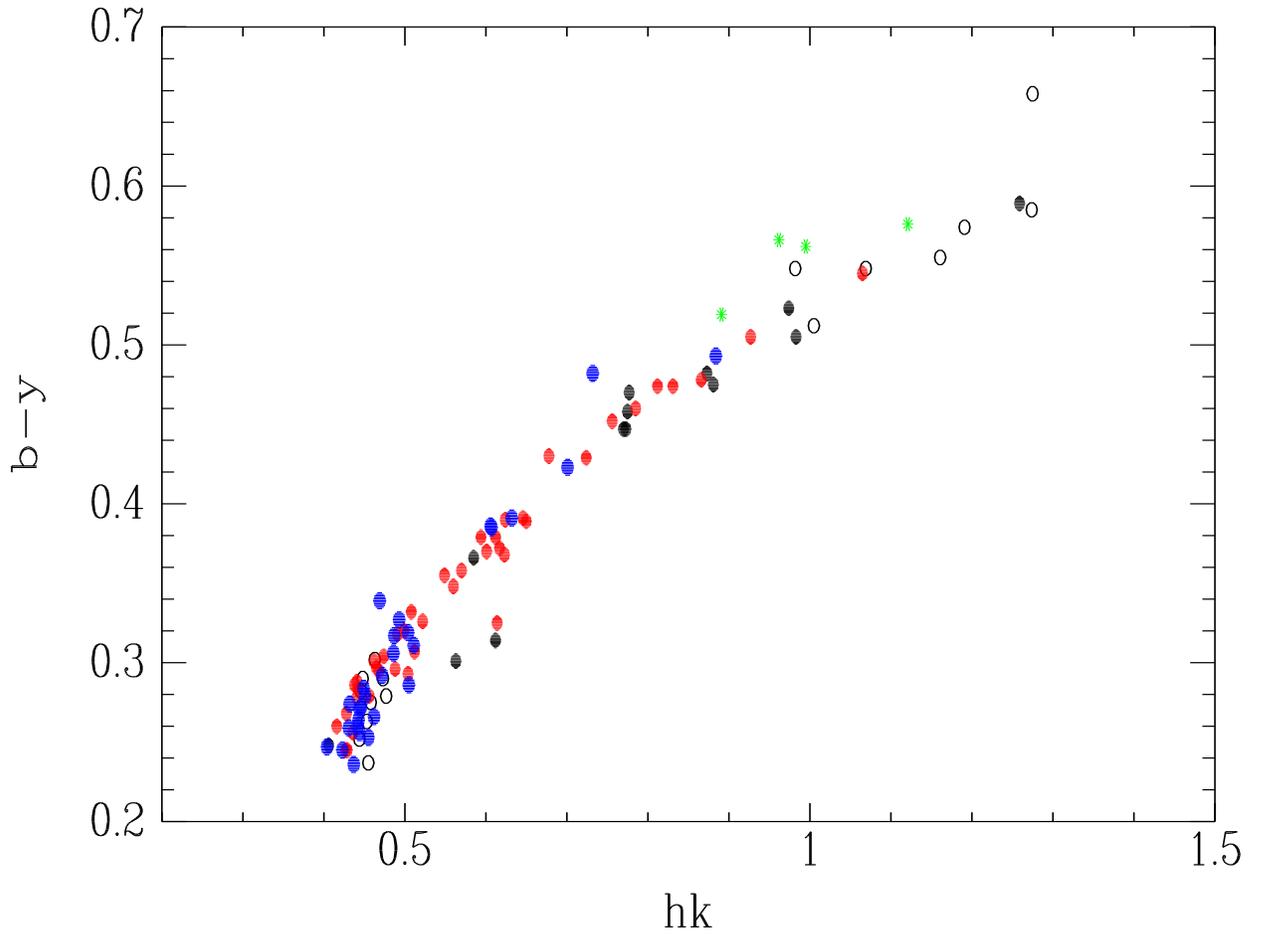}
\caption{Color-color relation for the non-red-giant stars of Fig. 6. Symbols have the same meaning as Fig. 5.}
\end{figure}

\clearpage
\begin{figure}
\includegraphics[width=\columnwidth,angle=270, scale=0.80]{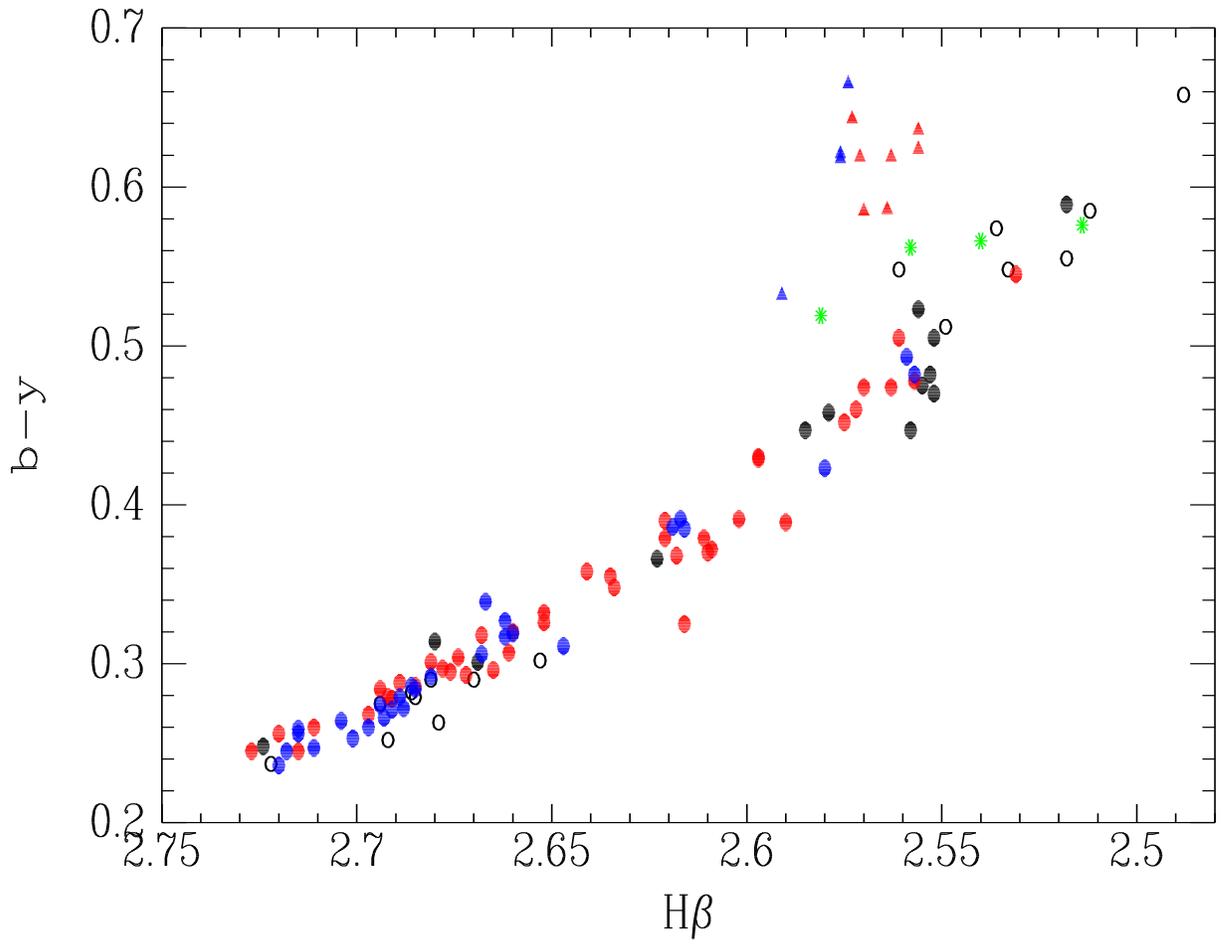}
\caption{$(b-y)$, H$\beta$ relation for possible members of NGC 752. Symbols have the same meaning as Fig. 5.}
\end{figure}

\clearpage
\begin{figure}
\includegraphics[width=\columnwidth,angle=0, scale=0.80]{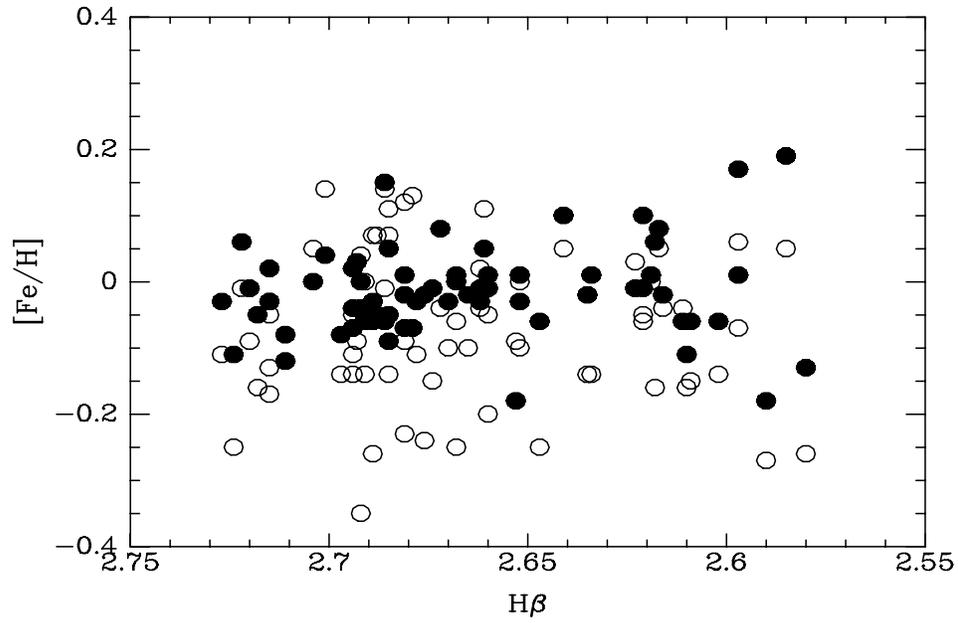}
\caption{[Fe/H] estimates from $m_1$ (open circles) and $hk$ (filled circles) for 68 dwarfs as a function of H$\beta$.}
\end{figure}

\clearpage
\begin{figure}
\includegraphics[width=\columnwidth,angle=270, scale=0.80]{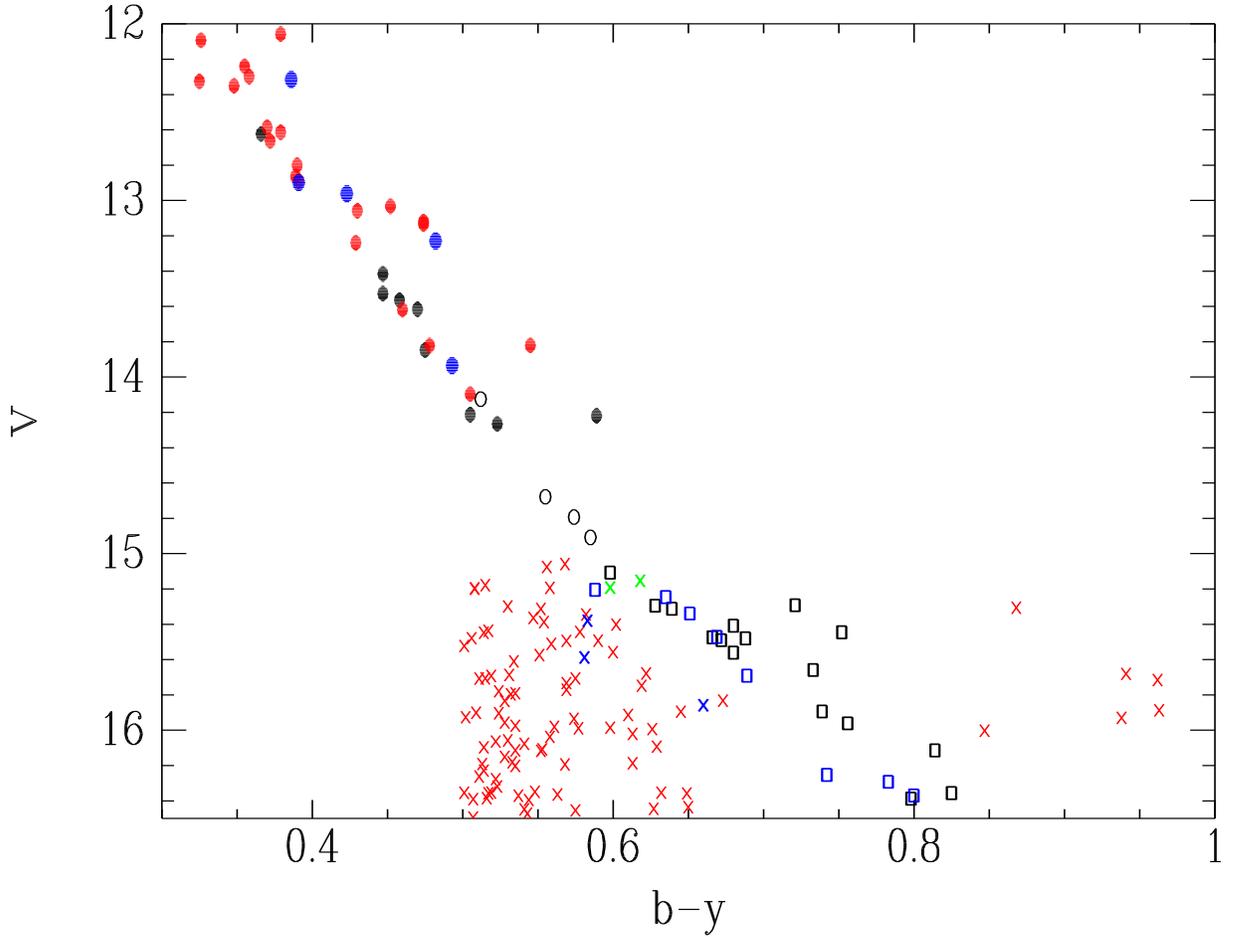}
\caption{Identification of photometric members in the CMD. Symbols have the same meaning as in Fig. 5, with the addition of open 
black squares (16 probable members from our data alone), open blue squares (members in common with \citet{BA11}), 
blue crosses (photometric members of \citet{BA11} which we classify as nonmembers), red crosses (nonmembers from our data alone),
 and green crosses (stars with indices which classify them as evolved stars). }
\end{figure}

\clearpage
\begin{figure}
\includegraphics[width=\columnwidth,angle=270, scale=0.80]{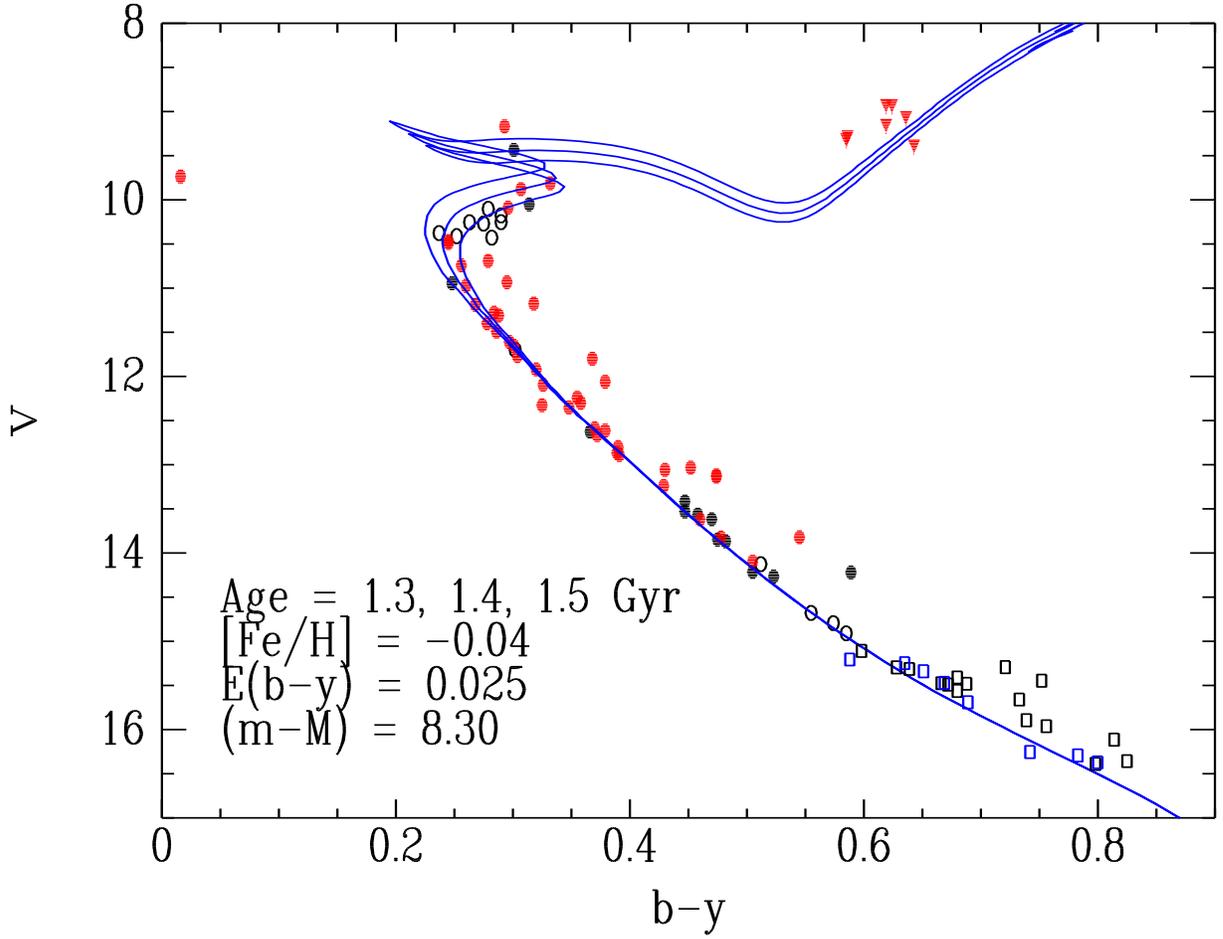}
\caption{CMD fit of Victoria-Regina scaled solar isochrones to probable single-star members of NGC 752. Symbols have the same meaning as
in Fig. 5 and Fig. 10. Isochrone ages are 1.3, 1.4, and 1.5 Gyr and have been adjusted to $E(b-y)$ = 0.025 and $(m-M)$ = 8.30.}
\end{figure}

\clearpage
\begin{deluxetable}{lccrrrc}
\thispagestyle{empty}
\tablecolumns{7}
\tablenum{1}
\tabletypesize\small
\tablewidth{0pc}
\tablecaption{$V$ magnitude Transformations}
\tablehead{
\colhead{Source} & \colhead{N}& \colhead{Color Index}& \colhead{A}& \colhead{B}& \colhead{Std.Dev.}& \colhead{Excluded Stars} }
\startdata
Johnson (1953)   		&  37 	&  $B-V$ 	&  -0.017  &     0.032 & 0.009 &  \cr
Eggen (1963)      		&  24 	&  $B-V$ 	&   0.051  &   -0.048  & 0.020 &  \cr
Jennens \& Helfer (1975) 	& 10 	&       &  0.000  &   -0.026  & 0.012 &  \cr
Twarog (1983)   		& 17  	&  $b-y$ 	&  -0.050  &    -0.002 & 0.008 &  \cr
Rufener (1981, 1988) 		&  39 	& 	&  0.000   &  -0.003   & 0.012 &  135, 201, 218, 273 \cr
Joner \& Taylor (1995) 		& 15 	& 	&  0.000   &   0.000   & 0.004 & \cr 
Barta\v{s}\=ui\.te et al. (2007)	&  37 	&  $Y-V$ 	&  -0.048  &     0.022 & 0.007 &  88 \cr
Anthony-Twarog \& Twarog (2007) &    20	&  	&   0.000  &    -0.024 & 0.008 & \cr\cr
\enddata
\tablecomments{Coefficients A and B characterize the relationship between standard and source $V$ magnitudes as follows:  
$V_{Standard} - V_{Source} = A (Color Index) + B$.}
\end{deluxetable}

\clearpage
\begin{deluxetable}{rrrrrrrrrrrrrc}
\thispagestyle{empty}
\tablecolumns{14}
\tablenum{2}
\tabletypesize\scriptsize
\tablewidth{0pc}
\tablecaption{Merged Photoelectric Secondary Standards in NGC 752}
\tablehead{
\colhead{H number } & \colhead{$V$ } &\colhead{sem} &\colhead{$b-y$} &\colhead{sem} &\colhead{$m_1$} &\colhead{sem} &
\colhead{$c_1$} & \colhead{sem} &\colhead{H$\beta$} &\colhead{sem} &\colhead{$hk$} &\colhead{sem} &\colhead{Comments} }
\startdata 
    27 &   9.148 &  0.004 & 0.627 & 0.009 & 0.368 & 0.013 & 0.408 & 0.015 & 0.000 & 0.000 & 1.091 & 0.012 \cr
    39 &   8.086 &  0.006 & 0.600 & 0.010 & 0.333 & 0.010 & 0.431 & 0.007 & 0.000 & 0.000 & 0.000 & 0.000 \cr
    58 &  10.489 &  0.007 & 0.260 & 0.001 & 0.149 & 0.008 & 0.621 & 0.011 & 2.707 & 0.008 & 0.445 & 0.033 & Var? \cr
    61 &  10.043 &  0.003 & 0.249 & 0.005 & 0.155 & 0.008 & 0.669 & 0.011 & 2.708 & 0.008 & 0.000 & 0.000 \cr
    62 &  11.209 &  0.007 & 0.268 & 0.005 & 0.147 & 0.008 & 0.541 & 0.011 & 2.702 & 0.010 & 0.000 & 0.000 \cr
    64 &  10.537 &  0.004 & 0.239 & 0.001 & 0.168 & 0.008 & 0.686 & 0.011 & 2.734 & 0.008 & 0.430 & 0.023 \cr
    66 &  10.923 &  0.003 & 0.293 & 0.002 & 0.150 & 0.000 & 0.461 & 0.001 & 2.672 & 0.011 & 0.488 & 0.023 \cr
    75 &   8.966 &  0.003 & 0.622 & 0.026 & 0.362 & 0.039 & 0.420 & 0.020 & 0.000 & 0.000 & 1.079 & 0.033 \cr
    77 &   9.372 &  0.003 & 0.640 & 0.004 & 0.376 & 0.005 & 0.441 & 0.017 & 0.000 & 0.000 & 1.098 & 0.012 \cr
    88 &  11.759 &  0.002 & 0.324 & 0.001 & 0.149 & 0.005 & 0.426 & 0.007 & 2.654 & 0.010 & 0.000 & 0.000 \cr
  & & & & & & & & & & & & & \cr
    96 &  10.377 &  0.001 & 0.241 & 0.005 & 0.167 & 0.008 & 0.717 & 0.011 & 2.722 & 0.010 & 0.000 & 0.000 \cr
   105 &  10.257 &  0.003 & 0.274 & 0.002 & 0.153 & 0.008 & 0.667 & 0.011 & 2.692 & 0.008 & 0.453 & 0.001 \cr
   106 &  10.501 &  0.005 & 0.244 & 0.006 & 0.162 & 0.008 & 0.659 & 0.007 & 2.714 & 0.007 & 0.000 & 0.000 \cr
   108 &   9.161 &  0.002 & 0.295 & 0.003 & 0.160 & 0.004 & 0.635 & 0.004 & 2.679 & 0.004 & 0.000 & 0.000 \cr
   110 &   8.956 &  0.003 & 0.531 & 0.005 & 0.266 & 0.002 & 0.457 & 0.012 & 0.000 & 0.000 & 0.815 & 0.029 \cr
   126 &  10.101 &  0.003 & 0.284 & 0.001 & 0.140 & 0.008 & 0.640 & 0.011 & 2.684 & 0.008 & 0.474 & 0.013 \cr
   129 &  10.895 &  0.005 & 0.246 & 0.005 & 0.151 & 0.010 & 0.634 & 0.007 & 2.705 & 0.007 & 0.000 & 0.000 & Var? \cr
   135 &  11.225 &  0.003 & 0.298 & 0.001 & 0.155 & 0.002 & 0.507 & 0.000 & 2.690 & 0.010 & 0.456 & 0.011 & Con? \cr
   139 &  11.757 &  0.004 & 0.312 & 0.005 & 0.136 & 0.009 & 0.455 & 0.003 & 2.672 & 0.007 & 0.473 & 0.012 \cr
   166 &   9.857 &  0.001 & 0.261 & 0.003 & 0.158 & 0.005 & 0.644 & 0.004 & 2.700 & 0.004 & 0.000 & 0.000 \cr
       & & & & & & & & & & & & & \cr
   171 &  10.189 &  0.002 & 0.284 & 0.002 & 0.156 & 0.001 & 0.625 & 0.009 & 2.679 & 0.001 & 0.000 & 0.000 \cr
   177 &  10.170 &  0.004 & 0.317 & 0.002 & 0.152 & 0.004 & 0.566 & 0.004 & 0.000 & 0.000 & 0.000 & 0.000 \cr
   187 &  10.437 &  0.003 & 0.275 & 0.004 & 0.155 & 0.004 & 0.554 & 0.002 & 2.694 & 0.001 & 0.441 & 0.024 \cr
   189 &  11.284 &  0.003 & 0.280 & 0.002 & 0.145 & 0.000 & 0.516 & 0.001 & 2.692 & 0.008 & 0.447 & 0.033 \cr
   192 &  10.746 &  0.001 & 0.253 & 0.006 & 0.000 & 0.000 & 0.000 & 0.000 & 0.000 & 0.000 & 0.473 & 0.030 \cr
   193 &  10.202 &  0.002 & 0.239 & 0.002 & 0.192 & 0.001 & 0.714 & 0.002 & 2.726 & 0.004 & 0.445 & 0.009 \cr
   196 &  10.254 &  0.002 & 0.276 & 0.002 & 0.159 & 0.004 & 0.630 & 0.004 & 2.694 & 0.009 & 0.000 & 0.000 \cr
   197 &  11.602 &  0.002 & 0.294 & 0.003 & 0.154 & 0.005 & 0.467 & 0.005 & 2.678 & 0.008 & 0.000 & 0.000 \cr
   205 &   9.899 &  0.004 & 0.279 & 0.002 & 0.155 & 0.001 & 0.559 & 0.002 & 2.690 & 0.001 & 0.436 & 0.022 \cr
   206 &  10.019 &  0.004 & 0.313 & 0.001 & 0.153 & 0.003 & 0.581 & 0.005 & 2.671 & 0.002 & 0.436 & 0.013 \cr
  & & & & & & & & & & & & & \cr
   208 &   8.950 &  0.003 & 0.669 & 0.002 & 0.413 & 0.005 & 0.381 & 0.005 & 2.566 & 0.008 & 1.120 & 0.006 \cr
   209 &   9.741 &  0.002 & 0.018 & 0.002 & 0.166 & 0.001 & 0.991 & 0.002 & 2.919 & 0.001 & 0.286 & 0.016 \cr
   213 &   9.030 &  0.003 & 0.621 & 0.001 & 0.367 & 0.004 & 0.407 & 0.008 & 0.000 & 0.000 & 1.072 & 0.022 \cr
   217 &  10.428 &  0.001 & 0.282 & 0.002 & 0.152 & 0.006 & 0.636 & 0.003 & 2.696 & 0.001 & 0.000 & 0.000 \cr
   218 &  10.078 &  0.003 & 0.300 & 0.005 & 0.145 & 0.008 & 0.603 & 0.011 & 2.681 & 0.008 & 0.000 & 0.000 \cr
   220 &   9.593 &  0.002 & 0.604 & 0.004 & 0.405 & 0.006 & 0.370 & 0.008 & 2.575 & 0.005 & 0.000 & 0.000 \cr
   222 &  10.966 &  0.001 & 0.268 & 0.003 & 0.151 & 0.003 & 0.610 & 0.006 & 2.713 & 0.005 & 0.391 & 0.001 \cr
   234 &  10.679 &  0.003 & 0.279 & 0.003 & 0.156 & 0.006 & 0.570 & 0.002 & 2.706 & 0.007 & 0.455 & 0.026 \cr
   238 &   9.960 &  0.004 & 0.305 & 0.002 & 0.141 & 0.008 & 0.608 & 0.011 & 2.666 & 0.008 & 0.465 & 0.015 \cr
   254 &  10.920 &  0.005 & 0.242 & 0.005 & 0.156 & 0.007 & 0.652 & 0.010 & 2.713 & 0.006 & 0.414 & 0.014 \cr
       & & & & & & & & & & & & & \cr
   259 &  11.393 &  0.006 & 0.277 & 0.006 & 0.147 & 0.007 & 0.522 & 0.010 & 2.696 & 0.006 & 0.393 & 0.014 \cr
   261 &  11.174 &  0.002 & 0.315 & 0.001 & 0.160 & 0.004 & 0.434 & 0.003 & 2.671 & 0.008 & 0.000 & 0.000 \cr
   263 &  10.948 &  0.003 & 0.240 & 0.001 & 0.160 & 0.006 & 0.628 & 0.007 & 2.712 & 0.002 & 0.000 & 0.000 \cr
   266 &  11.229 &  0.009 & 0.277 & 0.005 & 0.152 & 0.007 & 0.526 & 0.010 & 2.702 & 0.008 & 0.000 & 0.000 \cr
   295 &   9.292 &  0.003 & 0.584 & 0.005 & 0.350 & 0.012 & 0.385 & 0.003 & 2.570 & 0.010 & 1.063 & 0.023 \cr
   300 &   9.595 &  0.003 & 0.258 & 0.006 & 0.146 & 0.004 & 0.647 & 0.005 & 2.692 & 0.005 & 0.478 & 0.020 \cr
   311 &   9.054 &  0.002 & 0.641 & 0.009 & 0.384 & 0.014 & 0.423 & 0.019 & 2.566 & 0.010 & 0.000 & 0.000 \cr
\enddata
\end{deluxetable}

\clearpage
\begin{deluxetable}{ccrrrrc}
\thispagestyle{empty}
\tablecolumns{7}
\tablenum{3}
\tablewidth{0pc}
\tablecaption{Calibration Coefficients}
\tablehead{
\colhead{Index} & \colhead{Class} & \colhead{$a$} & \colhead{$b$} & \colhead{$c$} & \colhead{Num.} & \colhead{Residuals}  }
\startdata 
$V$     &   all  &  1.000  &  0.041  &  1.581    & 105 & 0.010  \\
$hk$    &  all  &  1.175  &  0.000  &  2.079  & 26  & 0.023 \\
H$\beta$ &  blue  &  1.17  &  0.000  &  0.308  & 35  & 0.008 \\
H$\beta$ &  red  &  1.12  &  0.000  &  0.411  & 38  & 0.008 \\
$b-y$   &   evolved blue & 1.067  &  0.000  &  0.243 &  46 &  0.005\\
$b-y$   &   red dwarf &  0.900  &  0.000  &  0.276  & 14  & 0.005\\
$m_1$   &   blue  & 1.000  &  0.000  & -0.954  & 35  & 0.008\\
$m_1$   &    evolved & 1.000  &  -0.454  & -0.839  & 24 & 0.014\\
$m_1$   &  red dwarf &  1.000 & 0.450  & -1.066   &  14  & 0.020\\
$c_1$   &   unevolved & 1.095  &   0.000  &   0.189  &   49  & 0.025\\
$c_1$   & evolved  & 1.000  &   0.338  &  0.148  &   25 &  0.031\\
\enddata
\tablecomments{For each index $X_i$, the calibrated value is $a_i X_i + b_i (b-y)_{instr} + c_i. $}
\end{deluxetable}

\clearpage
\begin{deluxetable}{rrrrrrrrrrrrrrrrrrrrrrrrrrrr}
\thispagestyle{empty}
\rotate
\tablecolumns{28}
\tablenum{4}
\tabletypesize\tiny
\tablewidth{0pc}
\setlength{\tabcolsep} {0.03in}
\tablecaption{Photometry in NGC 752}
\tablehead{
\colhead{$\alpha(2000)$} & \colhead{$\delta(2000)$} & \colhead{$V$} & \colhead{$b-y$} & \colhead{$m_1$} & \colhead{$c_1$} & \colhead{$hk$} & 
\colhead{H$\beta$} & \colhead{$\sigma_V$} & \colhead{$\sigma_{by}$} & \colhead{$\sigma_{m1}$} & \colhead{$\sigma_{c1}$} & \colhead{$\sigma_{hk}$} & \colhead{$\sigma_{\beta}$} & 
\colhead{$N_y$} & \colhead{$N_b$} & \colhead{$N_v$} & \colhead{$N_u$} & \colhead{$N_{Ca}$} & \colhead{$N_n$} & \colhead{$N_w$} & 
\colhead{WEBDA} & \colhead{$ID_{PL}$} & \colhead{$ID_H$} & \colhead{$ID_{RV}$} & \colhead{$ID_{ST}$} & \colhead{Memb.} & \colhead{Class}  }
\startdata 
  1.961863 &37.66953 & 7.119 & 0.741 & 0.535 & 0.394 & 1.335 & 9.999 &0.003& 0.008& 0.013 &0.012 &0.012 &9.999&  7 & 7& 17 &18 &34 & 0 & 0 & 215 & 882 & 215 &   1 & 285 &   0 &  G \\
  1.932012 &37.86701 & 8.085 & 0.596 & 0.342 & 0.417 & 1.049 & 2.572 &0.003& 0.006& 0.008 &0.006 &0.008 &0.006& 17 &15& 15 &17 &23 & 7 & 6 &  39 & 394 &  39 &  26 &  64 &   0 &  G \\
  1.980912 &37.68589 & 8.357 & 0.082 & 0.203 & 0.985 & 0.370 & 2.871 &0.006& 0.008& 0.016 &0.020 &0.012 &0.011&  5 & 5&  5 & 7 & 6 & 1 & 1 & 309 &1168 & 309 &  64 & 394 &   0 &  B \\
  1.927771 &37.88174 & 8.917 & 0.619 & 0.361 & 0.438 & 1.099 & 2.563 &0.002& 0.003& 0.004 &0.003 &0.004 &0.003&  7 & 7&  6 & 7 & 8 & 2 & 2 &  24 & 350 &  24 &  25 &  41 &  99 &  G \\
  1.950852 &38.13387 & 8.920 & 0.624 & 0.378 & 0.395 & 1.104 & 2.556 &0.002& 0.003& 0.004 &0.004 &0.003 &0.002& 14 &12& 15 &17 &17 & 6 & 5 & 137 & 687 & 137 &   0 & 202 &  99 &  G \\
  1.960438 &37.66030 & 8.940 & 0.665 & 0.397 & 0.390 & 1.147 & 2.574 &0.001& 0.002& 0.003 &0.003 &0.003 &0.003& 30 &29& 20 &20 &35 &18 &16 & 208 & 858 & 208 &   0 & 278 &  99 &  G \\
  1.947371 &38.03264 & 8.953 & 0.532 & 0.273 & 0.435 & 0.797 & 2.591 &0.002& 0.003& 0.004 &0.004 &0.004 &0.002& 19 &18& 19 &21 &23 & 9 & 9 & 110 & 630 & 110 &  33 & 174 &  95 &  G \\
  1.938685 &37.96680 & 8.967 & 0.618 & 0.362 & 0.432 & 1.100 & 2.576 &0.002& 0.003& 0.006 &0.007 &0.004 &0.008& 11 &12& 12 &12 &15 & 6 & 6 &  75 & 506 &  75 &  29 & 119 &  98 &  G \\
  1.960795 &37.76987 & 9.023 & 0.621 & 0.369 & 0.400 & 1.074 & 2.576 &0.002& 0.002& 0.003 &0.003 &0.003 &0.002& 40 &38& 34 &36 &49 &21 &19 & 213 & 867 & 213 &  60 & 283 &  99 &  G \\
  1.981216 &37.81636 & 9.054 & 0.636 & 0.398 & 0.385 & 1.116 & 2.556 &0.001& 0.002& 0.004 &0.004 &0.004 &0.003& 15 &16& 16 &18 &18 & 7 & 5 & 311 &1172 & 311 &  65 & 397 &  99 &  G \\ 
           &         & & & & & & & & & & & & & & & & & & & & & & & & & & \\
  1.928551 &37.63222 & 9.143 & 0.619 & 0.374 & 0.416 & 1.083 & 2.571 &0.004& 0.004& 0.006 &0.006 &0.005 &0.005&  8 & 8&  6 & 7 & 9 & 4 & 3 &  27 & 356 &  27 &  20 &  47 &  99 &  G \\
  1.971785 &37.80309 & 9.149 & 0.213 & 0.154 & 0.630 & 0.413 & 2.733 &0.003& 0.005& 0.008 &0.008 &0.007 &0.004& 31 &26& 21 &25 &30 &20 &15 & 271 &1033 & 271 &  62 & 349 &   0 &  B \\
  1.947187 &38.02248 & 9.167 & 0.293 & 0.150 & 0.660 & 0.504 & 2.672 &0.001& 0.003& 0.004 &0.005 &0.004 &0.004& 20 &20& 18 &21 &22 & 9 & 8 & 108 & 626 & 108 &  32 & 169 &  99 &  B \\
  1.924571 &37.99932 & 9.273 & 0.586 & 0.333 & 0.436 & 1.030 & 2.564 &0.002& 0.004& 0.006 &0.006 &0.006 &0.006&  9 & 8&  6 & 6 & 9 & 4 & 4 &  11 & 308 &  11 &   0 &  21 &  94 &  G \\
  1.974815 &37.86064 & 9.305 & 0.585 & 0.354 & 0.382 & 1.030 & 2.570 &0.003& 0.004& 0.004 &0.004 &0.005 &0.005& 24 &26& 22 &23 &30 &17 &15 & 295 &1089 & 295 &  67 & 377 &  99 &  G \\
  1.939415 &37.60233 & 9.376 & 0.643 & 0.375 & 0.441 & 1.116 & 2.573 &0.003& 0.004& 0.005 &0.004 &0.004 &0.004& 13 &14& 12 &12 &15 & 8 & 7 &  77 & 512 &  77 &  45 & 123 &  98 &  G \\
  1.953353 &37.98999 & 9.434 & 0.301 & 0.170 & 0.594 & 0.563 & 2.669 &0.002& 0.002& 0.004 &0.004 &0.004 &0.004& 24 &22& 24 &26 &30 &16 &13 & 159 & 728 & 159 &   0 & 225 &  99 &  B \\
  1.962645 &37.65648 & 9.571 & 0.607 & 0.391 & 0.406 & 1.069 & 2.578 &0.002& 0.003& 0.004 &0.005 &0.005 &0.004& 34 &34& 20 &20 &35 &21 &20 & 220 & 895 & 220 &   0 & 290 &   0 &  G \\
  1.976816 &37.75319 & 9.595 & 0.272 & 0.159 & 0.608 & 0.446 & 2.688 &0.003& 0.004& 0.005 &0.005 &0.005 &0.003& 23 &22& 21 &22 &24 &11 &10 & 300 &1117 & 300 &  63 & 382 &  99 &  B \\
  1.932206 &38.04229 & 9.698 & 0.995 & 0.769 & 0.120 & 2.030 & 2.574 &0.002& 0.003& 0.007 &0.010 &0.005 &0.002& 15 &15& 12 &12 &15 & 8 & 8 &  40 & 398 &  40 & 474 &  65 &   0 &  G \\
\enddata
\tablecomments{In addition to identifications from the WEBDA database, identifications from Platais (1991), Heinemann (1926), Rohlfs \& Vanysek (1961) and Stock (1985) are included.  The final columns denote the membership probability from Platais (1991) and the calibration equation class.}
\end{deluxetable}

\end{document}